\documentclass[10pt|11pt|12pt]{article}
\usepackage{cite}
\usepackage{amssymb,amsmath,latexsym,mathrsfs}
\usepackage{psfrag}
\usepackage{bm}
\usepackage{cite}
\usepackage{url}
\usepackage{color}
\usepackage{comment}
\usepackage{graphicx}
\usepackage{caption}
\usepackage{subcaption}
\usepackage{booktabs}
\usepackage{multirow} 
\usepackage{mathtools} 
\usepackage{extarrows} 
\usepackage{centernot}

\usepackage{authblk}

\usepackage{cite}
\usepackage{amsthm}

\newtheorem{theorem}{Theorem}[section]

\newtheorem{lemma}[theorem]{Lemma}

\newtheorem{corollary}[theorem]{Corollary}

\newtheorem{compsol*}{Complete Solution}

\usepackage{datetime}

\date{\displaydate{date}}

\usepackage[many]{tcolorbox}
\usetikzlibrary{calc}

\definecolor{myblue}{RGB}{0,163,243}

\tcbset{mystyle/.style={
  breakable,
  enhanced,
  outer arc=0pt,
  arc=0pt,
  colframe=myblue,
  colback=myblue!20,
  attach boxed title to top left,
  boxed title style={
    colback=myblue,
    outer arc=0pt,
    arc=0pt,
    top=3pt,
    bottom=3pt,
    },
  fonttitle=\sffamily
  }
}

\newtcolorbox[auto counter,number within=section]{bluebox}[1][]{
  mystyle,
  title=\textbf{\large Significance},
  overlay unbroken and first={
      \path
        let
        \p1=(title.north east),
        \p2=(frame.north east)
        in
        node[anchor=west,font=\sffamily,color=myblue,text width=\x2-\x1] 
        at (title.east) {#1};
  }
}

\begin{document}

\title{\LARGE \bf  A new threshold reveals
the uncertainty about the effect of school opening
on diffusion of Covid-19.}


\author[1]{Alberto Gandolfi}
\author[2]{Andrea Aspri}
\author[3]{Elena Beretta}
\author[1]{Khola Jamshad}
\author[1]{Muyan Jiang}
\affil[1]{New York University Abu Dhabi}
\affil[2]{Università di Pavia}
\affil[3]{New York University Abu Dhabi
and Politecnico di Milano}



\vspace{5mm}

\date{}

\maketitle

\thispagestyle{plain}
\pagestyle{plain}

\let\thefootnote\relax\footnotetext{
AMS 2020 subject classifications: 
37N99

Key words and phrases: epidemic, Covid-19, effect of school opening, in-class teaching, remote learning, SIR model, subpopulations, phase transition,  vaccines.

\thanks{  }

}

\begin{abstract}
{ \bf 
We aim at clarifying the controversy about
the effects of school openings or closures
on the course of the Covid-19 pandemic.

The mathematical analysis of compartmental models
with subpopulations shows that   the in-school contact
rates  affects the overall course of the
pandemic only above a certain
threshold that separates an influence phase
from a non-influence one.
The threshold, that we calculate via
linear approximation in several cases,
seems to appear in all contexts, including
outbreaks or new
strains upsurge,  lockdowns,
and vaccination campaigns excluding children, albeit 
with different values.
Our theoretical findings are then confirmed by 
several data driven studies that have 
previously identified the phase transition
in specific cases.

Specific outcomes of this study are:
\begin{itemize}
    \item 
opposite conclusions reached by studies
of the same or
similar situations might depend on, possibly
small, differences in modeling or in parameter estimation
from the very noisy Covid-19 data, that result
in identifying different phases;
 \item  it is possible to keep schools open at any stage
of the Covid-19 pandemic, but suitably strict rules must be 
applied at all times or else this becomes highly
detrimental to virus containment efforts;
 \item as the threshold during vaccination turns out
 to correspond 
to the internal transmission rate that would 
lead to virus extinction
if the school population was isolated, the needed 
strict control can be sustained only for very brief
periods; as a result, either schools will have to 
face a prolonged closure or
children need to be vaccinated as well.
\end{itemize}

}

\end{abstract}

\vskip 10truemm


  \section{Introduction}
  This paper stems from the attempt that
  our group made some months ago to 
use some of the available data to clarify
the potential effects of school opening 
on the course of the Covid-19 outbreaks.
The question of opening or closing schools
had already turned out to be 
one of the most debated issues of the pandemic,
affecting more than one billion students, 
their teachers and communities.
  Schools had been closed at the early stages of the pandemic in 
almost every country, with classes held online for most of last year,  \cite{WHO, Viner2020};
but the, sometimes hurriedly arranged, remote teaching 
 was 
creating great difficulties to students, teachers and families \cite{Cao2020, Singh2020}.

As we kept exploring and trying to model data, we
ran into a seemingly very unstable situation.
Our compartmental models, similar to many others
currently used, were getting good approximations,
but they relied on parameters, such as
duration of clinical symptoms or  age 
dependent susceptibility, previously estimated by
other studies. We noticed then that the  effect of school opening that the models were  asserting 
depended in a crucial way on small changes 
in these parameters. We kept oscillating between
making the schools the culprits of the pandemic, 
or completely absolving them, a situation 
schematically represented in Figures \ref{20}-\ref{22}
below. When we looked at other studies appearing
at the time, it seemed that 
they also offered different, and sometimes opposite
views \cite{GriffithS_2(0)20,Walsh et al. 21,Lee et al.,Brauner et al,Keskinocak et al,Gandini et al
}, even with almost the same data.

Our current aim is to propose a reason for such
instability, and for the consequent
lack of reliability of many studies on this topic.
In addition, we would like to offer 
a theoretical framework on which school opening
policies could be based from now on.

Before starting, let's clarify 
that, in focusing on the effects of
in-school transmission on the course
of the pandemic, we
disregard many other relevant epidemiological issues 
concerning school opening, 
such as the possible role and availability  of teachers \cite{Gold et al.}; the sustainability of
school opening \cite{gandolfiI_2(0)};
as well as psychological, cultural, educational
aspects; that need then to be added when planning
concrete policies.

On the other hand, keeping models simple 
in the spirit of \cite{Bertozzi et al,
Roda et al.}, might help highlighting the fundamental
mechanisms at work. We consider, then, an SIR model with
two subpopulations and a minimal number of parameters,
studying how changes in the transmission rate
in one population affect the overall course of the infection.

The outcome of our analysis has been the existence of a
perhaps surprising threshold, below which
further reduction of in-school transmission, 
with closure for instance, has almost
no effect, but above which school opening
becomes the leading factor driving infections.

We proceeded to test
various scenarios: uncontrolled outbreaks or new 
strains upsurge,
which occur at the early stages of virus diffusion and after the release of successful
 rigid controls; lockdowns; and
 successful vaccination campaigns on the largest,
 out-of-school,
  subpopulation. And we later
  tested some more elaborate models
with pre- and a-symptomatic and many other
realistic features. We also review many data based
studies from the perspective of this
newly discovered phenomenon.
The sharp transition
showed up in each context, indicating that 
a general mechanisms is at work.

We proceed to describe the details of 
the phase transition phenomena, adding rigorous proofs
based on the linear approximation of 
the simplest models. We also compute the explicit
values of the thresholds in the various scenarios.

Of particular interest is the case of
a vaccination campaign of adults, in which 
the threshold for in-school transmission rates
turns out to be extremely small, giving a strong
indication of the need of a children
vaccination campaign.

 \section{Results}
 \vskip 0.5cm
 
\subsection {There is a phase transition in the effect of school transmission rates on the overall epidemic course.}
\label{sub1}
 In a simple SIR model with subpopulations $1$ and $2$,
 in-school and out-of-school,
 with fractions $S_1(0)$ and $S_2(0)$ of susceptibles, 
 respectively, 
 and 
 transmission rates $\beta_{ij}$ from population $j$ to population $i$, the effect of 
 $\beta_{11}$ on the course of the epidemic
 undergoes a phase transition
 with threshold 
 \begin{equation}
 \beta_{11}^*=
 \beta_{22} S_2(0)/S_1(0).
 \end{equation}
 The
  total number of active cases is almost
  constant for all $\beta_{11} \in
  [0,\beta_{11}^*]$, and has a sharp increase
  for $\beta_{11} > \beta_{11}^*$. 
 An effective containment of the
  effect of a change of $\beta_{11} $ is achieved if
 \begin{equation} \label{FirstCriticalPoint}
 \beta_{11} < \beta_{22} S_2(0)/S_1(0)- \alpha \sqrt{\beta_{12} \beta_{21} S_2(0)/S_1(0)},
 \end{equation}
 for some  suitable multiple
 $\alpha$, while there is a substantial effect if 
 \begin{equation} \label{FirstCriticalPoint2}
 \beta_{11} > \beta_{22} S_2(0)/S_1(0)+ \alpha \sqrt{\beta_{12} \beta_{21} S_2(0)/S_1(0)}.
 \end{equation}
Since in concrete cases $\beta_{12}, \beta_{21} <<\beta_{22} $ \cite{PCJ17,Kim2020} the above values are close to $
\beta_{11}^*$.

Calculations are done in a linear approximation of
the SIR model, which applies to the Covid-19 pandemic 
as the numbers of active cases
are kept relatively low by containment measures in the
early stages of the outbreaks \cite{Flaxman2020}.
In the linear approximation it is possible to formally
compute the total number of cases up to a certain
time $\overline t$, which corresponds to when a
 lock down is imposed. 
If the target is to contain the increase in the 
number of total cases up to $\overline t$ 
to a given percentage 
$\epsilon$, 
an explicit formula allows to compute
the maximal allowed value of $\beta_{11}$.

 In a realistic example with total population 
 and recovery rate $\gamma$ normalized to $1$,
 setting  $S_1(0)=0.2, S_2(0)\approx 0.8,  \beta_{12}= \beta_{21} =0.5, \beta_{22}=2 $, the critical point is
 $\beta_{11}^*\approx 8$.
 Assuming an initial fraction of $3 \times 10^{-5}$ of active cases in Subpopulation $2$ and none in Subpopulation $1$, a 
 rescaled time
 frame of $\hat t=5$ (corresponding to approximately
 $50$ days), and $\epsilon=0.3$, a suitable value of 
 $\alpha$
gives $\beta_{11} \leq 6.344$.

The first part of Figure \ref{Recap},
for $t \in [0,5]$, shows a simulation of the active cases
with the above values. One can  see that
school opening has  a moderate effect for small
values of  $\beta_{11} $, and then
the effect becomes dramatic as the values increase
past the critical point. More detailed illustrations
are in the online material.


It follows that closing schools, i.e. setting $\beta_{11}=0$, is of limited impact if the reproduction rate $\beta_{11} S_1(0)$ in school
is somewhat lower than $\beta_{22} S_2(0)$, the external reproduction rate, and 
of substantial impact otherwise. This provides harmless school opening options,
assuming that one has access to the reproduction rates in the subpopulations.

 \vskip 1cm

\subsection{The  phase transition is preserved under lock-down, 
 albeit with a different critical point.} \label{PhaseTransLock}
 An 
 analogous effect takes place when a lockdown is imposed.  
 If at some time $\overline{t}$ transmission rates are
 reduced to values $\overline \beta_{ij}$,
 corresponding to a subcritical reproduction number,
 then the effect of $\overline \beta_{11}$
 on the total number of active cases undergoes the same
 phase transition as during the outbreak,
 but with critical point 
 \begin{eqnarray}
\overline \beta_{11}^{*}=\frac{1}{S_{1}(\overline t)}
-\frac{\overline\beta_{12} \overline\beta_{21} S_{1}(\overline t) S_{2}(\overline t)}{ S_{1}(\overline t) (1 - \overline\beta_{22} S_{2}(\overline t))}.
\end{eqnarray}
More precisely,
let
\[
A=\frac{\Delta S (\overline\beta_{11})}{(S_1(\overline t)+
S_2(\overline t))}
=\frac{(S_1(\infty) -S_1(\overline t)+ S_2(\infty)- S_2(\overline t))}{(S_1(\overline t)+
S_2(\overline t))}
\]
indicate  the attack rate of the epidemic,
i.e.  the fraction of the initially susceptible population
 that is eventually infected 
by the disease in the course of the epidemic
from $\overline t$ to complete eradication.
It turns out that 
 a sufficient condition to ensure that $
\Delta S (\overline\beta_{11})$
 does not exceeds
$(1+\epsilon )\Delta S (0)$ is
\begin{eqnarray}
\overline\beta_{11}<  F(\epsilon)
\end{eqnarray}
where $F$ (see \eqref{CapitalF} below) is a function that depends on the
proportions of active cases and susceptible
individuals at 
time $\overline t$.

In a realistic example continuing the one 
for the outbreak, with $\overline\beta_{12}= \overline\beta_{21} =0.25, \overline\beta_{22}=1, \epsilon=0.3 $,
in order to contain the increase in attack rate
to no more than $30\%$ one needs now to have
\[\overline\beta_{11} < 2.9944.\] 
Although in a different scenario, this is smaller
than the value $ 6.344$ found in the outbreak,
as there the aim was just to avoid producing an even
more extended diffusion of the infection.

 The second part of Figure \ref{Recap}, 
 for $t \in [5,18]$, illustrates
 active cases in the lockdown scenario, with the above values of
 the model parameters. 
 
 \bigskip

When considering a complete outbreak-lockdown cycle,
the attack rate undergoes a similar 
transition, depending on the  values of 
the two transmission rates $\beta_{11}$ and $\overline \beta_{11} $.
If the pair is sufficiently closed
to $(0,0)$, then there is little change 
in $\Delta(S)$, while there is a drastic change for 
larger values of the two transmission rates
(see Figure \ref{PhaseTransition3}).

\vskip 1cm

\subsection{ Success of widespread vaccination of non-schooling individuals
requires internal reproduction number in schools to be
subcritical.} \label{sub3}
If a vaccination campaign for not-in-school individuals
is carried out, the total number of cases from
a restart of the epidemic to the complete disappearance
due to vaccination undergoes an analogous phase transition,
with threshold
\begin{eqnarray}
\widetilde \beta_{11}^{*}=
1/{S_1(0)}.
\end{eqnarray}
The attack rate
is only moderately changed for 
$\widetilde \beta_{11}$ below the threshold, while
the outcome of the vaccination process is 
substantially disrupted for 
larger values of $\widetilde \beta_{11}$.

Notice that if $\widetilde \beta_{11}<\widetilde \beta_{11}^{*}$
then the in-school reproduction number
$R_S=\widetilde \beta_{11} S_1(0) $
is subcritical, i.e. less than one.

With the data of the previous examples, suppose 
a vaccination program is introduced targeted
to a $60\%$ coverage in about $3$ months (Israel
kept this pace at the time we are writing, with 
schools almost completely closed). Since
the internal reproduction number is
$R_S=0.198872 \widetilde \beta_{11}$, then the critical
value for 
$\widetilde \beta_{11} $ is $5.02836$. This is seen in
the third part of Figure
\ref{Recap}, for $t\in [18,31]$.
To achieve a sensible containment 
that limits the number of extra infections to no more
than $30\%$ one needs to take $\widetilde{\beta}_{11}<3.03111$.


  \vskip 2cm

\subsection{From the point of view of
containing the epidemic, schools can be kept open at all times, 
with strict control measures}
Taken together, 
the  results obtained from simple
SIR models with subpopulations
show that, although
the values of the critical points are different, opening
of schools would not seriously affect the 
course of the pandemic at all times, provided
the internal transmission rate is kept low enough.
On the other hand, if the control is released,
then the effect of school opening becomes
dramatic.

\begin{figure}[ht]
    \centering
    \includegraphics[scale=0.35]{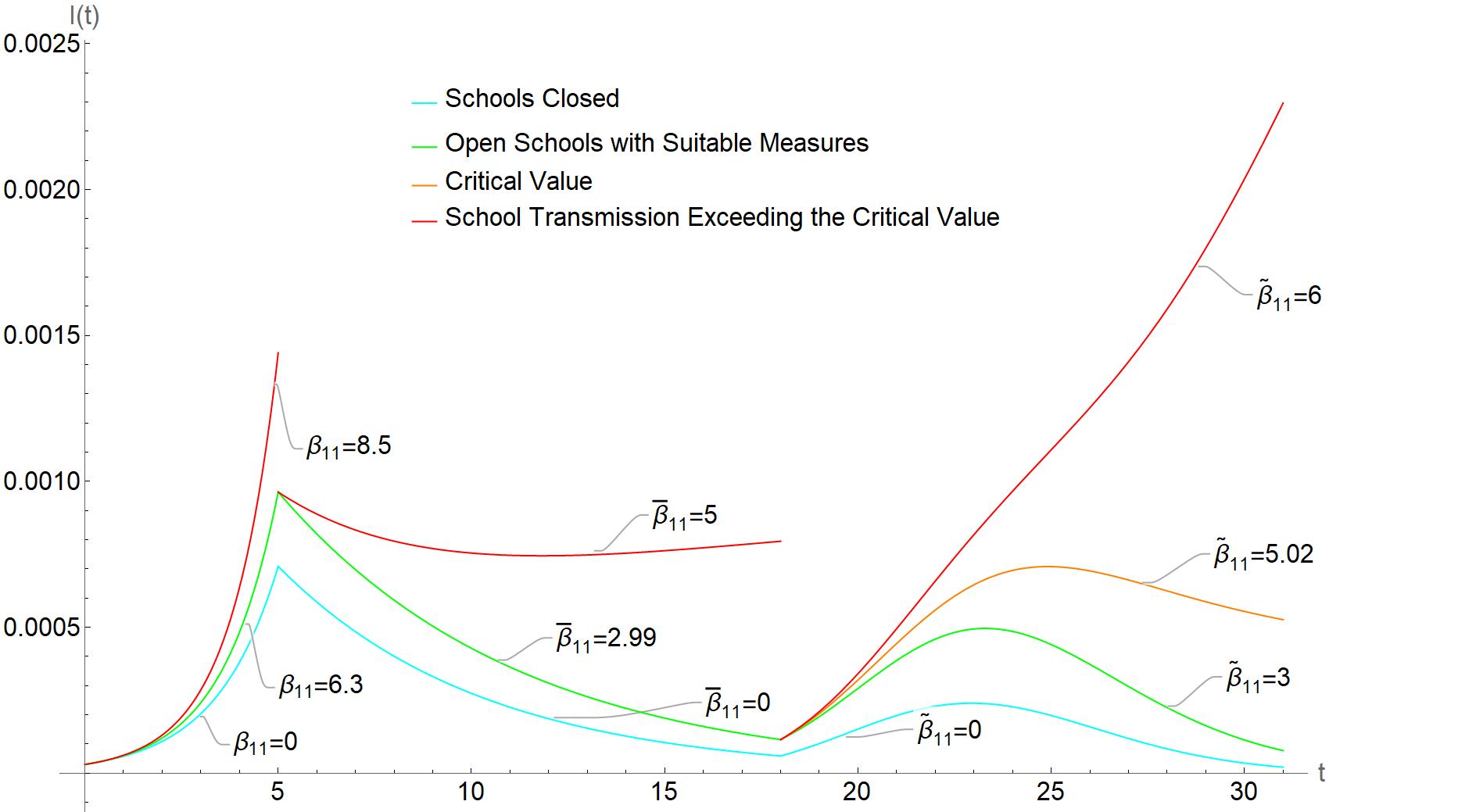}
    \caption{Daily active cases $I_1(t)+I_2(t)$  for various scenarios: 
    outbreak or new strain upsurge, lockdown, and vaccination.
    In each case, there is critical value for the
    in-school transmission rate
    $\beta_{11}$.
    Cyan curve is with closed schools, green
    for safe opening, orange for critical values, 
    red for values above criticality.
    }
     \label{Recap}
\end{figure}

Figure \ref{Recap} summarizes the numbers of 
active cases in the three scenarios we
have analyzed: the cyan curve corresponds to 
closed schools, while the green one is a subcritical 
pattern; the red curves, instead, show the risk that
the pandemic
spirals out of control because of insufficiently
controlled school opening.

 \vskip 1cm

 An analogous behavior takes place in more elaborate and realistic models, involving presymptomatic, asymptomatic,
 testing, isolation etc. Critical values
 appear for the in-school transmission rate, below
 which the effect of school opening on the
 epidemic trajectory is extremely contained.
 We  provide simulations in Section \ref{SPIAR}
 (see Figure \ref{fig:SPIAR}),
 and evidence from case studies here below.

 \vskip 1cm

\subsection{Evidence of phase
transition appears in several data driven case studies}
\label{EvidenceData}

The effect of a
phase transition seems to appear in all
data driven studies (see Section \ref{CaseStudies}). In most studies the
effect can be seen as the research
reaches a definite conclusion: in
some cases,  the data or  the model after
calibration correspond to a subcritical regime,
so that the study ends up asserting the
almost irrelevance of school opening on the pattern
of the epidemic for all the analyzed cases; in other
cases, the study determines a
supercritical setting, and then comes
to the opposite conclusion.

\begin{figure}
    \centering
    \includegraphics[scale=0.4]{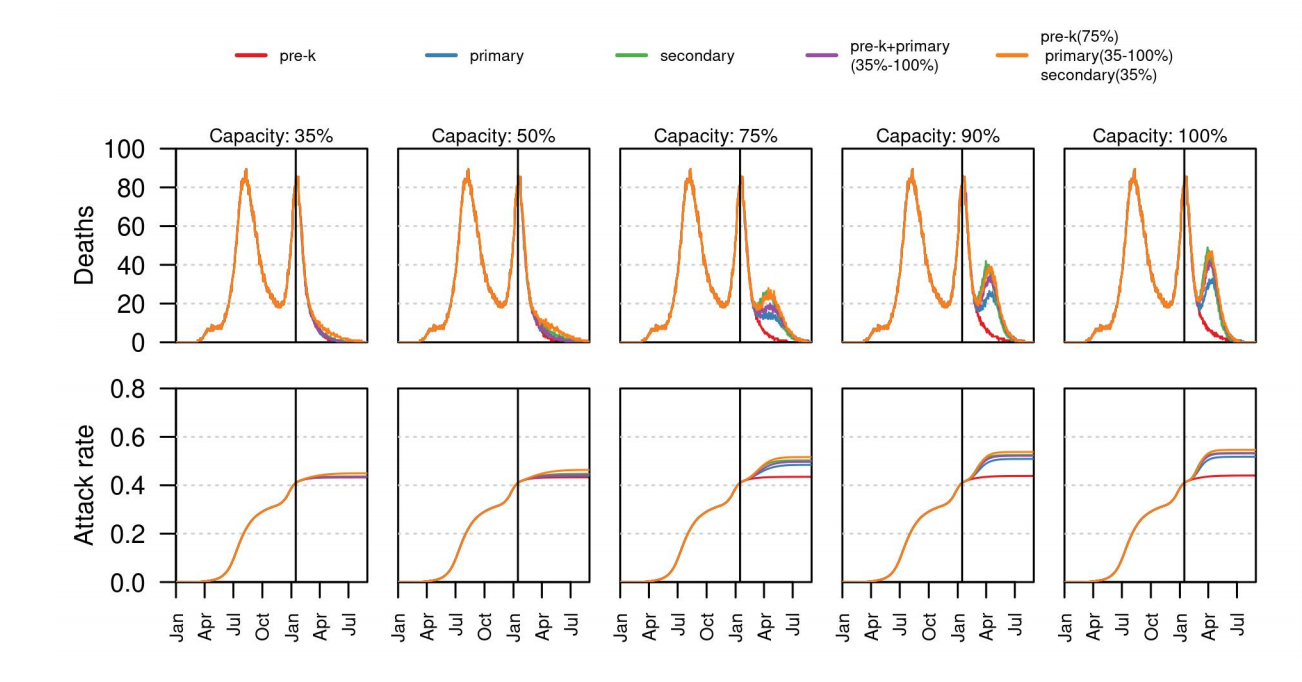}
    \caption{The impact of school reopening strategies in time
    as simulated in  \cite{Espana2021}
    from data from Bogot\`a, Colombia, for
    various values of the capacity, i.e. 
    the percentage of students allowed back at school.
     Each column shows a different
capacity level. Top panel shows the median daily incidence of deaths for each reopening strategy
based on grades. Bottom panel shows the estimated attack rate for each of the reopening
scenarios. Vertical black line shows the timing of school reopening (January 25, 2021). All
scenarios were simulated up to August 31, 2021. 
}
    \label{fig:espana}
\end{figure}

Some works include one or more parameters
that can be modulated to envision the effect of school reopening.
In these cases, one can see the effect of a
sharp transition from a subcritical, acceptable reopening, 
to an excessively impactful one. 
In \cite{Espana2021}, Figures $4$ and $5$,
for instance, one
can see that
up to $50\%$ capacity the effect of opening
schools is almost negligible, while it
becomes substantial above $75\%$; this is a likely
indication of
a critical point between these values.
For convenience, Figure $4$ in \cite{Espana2021}
is reproduced here in Figure \ref{fig:espana}.

A very detailed study of school opening in The Netherlands is conducted in
\cite{Rozhnova et. al}, and their conclusions are a clear description
of the phase transition. Using a data driven, elaborate
model, \cite{Rozhnova et. al} claims
that
their "analyses suggest that the impact of measures reducing school-based contacts depends on the remaining opportunities to reduce non-school-based contacts. If opportunities to reduce the effective reproduction number ($R_e$) with non-school-based measures are exhausted or undesired and $R_e$ is still close to $1$, the additional benefit of school-based measures may be considerable, particularly among older school children."
The first scenario of \cite{Rozhnova et. al} corresponds to
a subcritical in-school transmission rate, so that 
the effect of closing schools would be very moderate.
The second scenario seems to correspond to an in-school transmission
rate around the critical value, so that
both containment, in- or out-of-school, are effective.

\cite{Yuan2021} uses a detailed compartmental model and data from
the second semester 2020 in Toronto, 
an outbreak context, to estimate the 
likely impact of school opening; with parameters estimated
and collected from literature, the paper finds that 
opening school has little effect on the overall course
of the pandemic: in all scenarios presented in Figures $2$ and $4$
the difference between school opening and closure
is extremely contained. The findings of the 
research is then consistent with our 
phase transition scenario. In particular, \cite{Yuan2021}
finds that 
"school reopening was not the key driver in virus resurgence, but rather it was community spread that determined the outbreak trajectory"; in other words, the parameters of the models,
although not explicitly given in the paper,
are such that 
 the external transmission is preponderant. 
 As an additional finding, it is observed in
 \cite{Yuan2021} that, according to their model,
 "brief school closures did reduce infections when transmission risk within the home was low": in this case, a reduced transmission
 rate outside makes the one in school likely supercritical.

 \section{Confounding effects on 
 retrospective studies and
 forecasting}
 The presence of a threshold raises the specter of
 possible close to chaotic behaviors, a phenomenon 
 affecting epidemic forecasts as outlined in
 \cite{Castro et al.}. 
 Even with close-by initial data, if
 one of two systems crosses a critical threshold, then
 the future evolution might become extremely different
 from that of the other.
 For this reason, predictions on the future effects
 of in-school activities are going to be highly
 unreliable.
 
To make things worse, even retrospective studies
trying to evaluate the role of school openings or 
closures on the evolution of the pandemic 
run the risk of being completely untrustworthy.
Covid-19 data are affected by enormous errors, due
to the presence of asymptomatic, lack 
and partial reliability of testing,
difficulty in assessing close contacts etc.
It follows that estimation of parameters for
both statistical and model based studies are
affected by large errors.
In the presence of a threshold, even small errors
can lead to incorrect attribution of the situation
under observation to one phase, or
to conflicting attributions to two opposite phases
by different studies.
In such scenario, a retrospective study could 
misclassify the effect of school opening or closure;
and different studies even based on almost the same
data might end up reaching opposite conclusions.

\begin{figure}
    \centering
\includegraphics[scale=0.4]{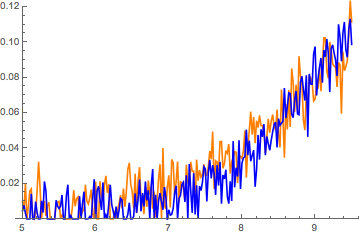}
    \caption{Synthetic data of daily active cases
    in a  two models, one with subcritical and  one
    with supercritical in-school transmission rates, 
    and slightly different initial number of cases. Gaussian
    noise has been added to make the example more realistic.}

    \label{20}
\end{figure}

In the noisy, synthetic data in Figure \ref{20},
the number of daily infected in a population
have been generated
with the same parameters, except that
\begin{eqnarray}
\text{ first model: } && \beta_{11} = 10, \beta_{2,2}=2, I_2(0)= 3 \times 10^{-5}\\
\text{ second model: } &&\beta_{11}= 6, \beta_{2,2}=2.57, I_2(0)= 5.5 \times 10^{-6}.
\end{eqnarray}
In school transmission rates are supercritical 
in the first case, and subcritical in the second.
But the different number of initial cases, a
value that is subject to errors of various order of 
magnitudes and is quite arbitrarily assigned in
the various studies, makes the two trajectory
basically indistinguishable.

\begin{figure}
    \centering
    \includegraphics[scale=0.4]{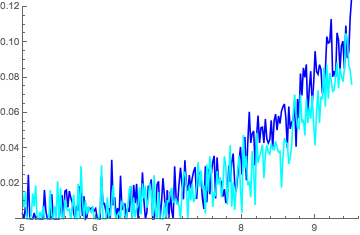}
    \caption{Reduction in daily cases due to school closure
     in the first scenario}
    
    \label{21}
\end{figure}

\begin{figure}
    \centering
    \includegraphics[scale=0.4]{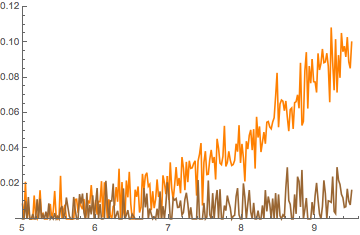}
    \caption{Reduction in daily cases
    due to school closure in the second scenario}
     \label{22}
\end{figure}

 In a retrospective study one is forced to 
 assign an initial value to the number of infected,
 and then estimate other parameters from the 
 observations. Both scenarios are then plausible, 
 depending on the chosen initial values.
 As Figure \ref{21} confirms, the research would 
 conclude in the first scenario, that closing 
 schools would have been basically useless.
 In the second scenario, however, the opposite
 conclusion would be drawn, as illustrated 
 by Figure \ref{22}.

\newpage
 
 \section{Conclusions}

Our results indicate that 
from the point of view
of the effect of school opening on the overall course 
of the pandemic, there are no fundamental reasons to keep schools closed.

However, school opening comes at a price: 
the  transmission rate in 
schools must be kept below a certain threshold 
that depends on the situation, and might not be 
easily determinable nor achievable.

The presence of a threshold is the likely cause
of the opposing views that many studies have
presented, between asserting almost irrelevance
of school opening, and others pointing to 
its significant effects. Minimal changes
in the overall conditions, or in
values of the estimated parameters 
may determine one phase of the other; 
this may result in  different attributions
of responsibility to school opening, 
and creates the possibility of  
an arbitrary identification of the phase
due to parameter estimation in the presence
of very noisy data.

In particular,
in the presence of a vaccination 
being carried out largely for out-of-school individuals
only,
there is a threshold below which schools can still 
be opened; it corresponds, however, to an 
internal reproduction number that would
eradicate the virus if
schools were completely isolated.
As this is highly demanding, such a containment 
seems to be sustainable for brief periods only,
after which vaccination for children
becomes the only viable possibility 
to return to normality.

\section{Limitations and related works}

Although the presence of a phase transition in the effect 
of the school transmission rate in the overall
course of the epidemic seems to have been
unnoticed so far in the literature, there are
many works related to ours.

In the direction opposite to the one taken here,
namely focusing on the sustainability
of opening from the point of containing the
number of cases of a single school, one can 
see \cite{gandolfiI_2(0)}.

Compartmental models with two subpopulations are
discussed in many works in general terms, see for instance a review in \cite{Kim2020}; and then applied to
the school opening issue in data driven analyses 
in \cite{ Yuan2021,
Domenico2021, Espana2021,Ismail2020}: we 
discuss the relation of some these results
with our work in Section  \ref{CaseStudies}.

Finally, other papers 
 \cite{Iwata2020, Ismail2020, Larosa2020}  make a purely statistical
evaluation of the effect of school opening (see Section  \ref{CaseStudies}).

\vskip 0.5cm
Our work has several limitations.
Our results are based on abstract, simplified models,
and, although they seem to be stable when more detailed
features are included, we cannot ensure that
they always take place in more complex models.

Even when a critical value can be estimated, insuring that the transmission rates are below their
relative thresholds is clearly a matter of
 distancing, testing, and other measures
 (see [https://science.sciencemag.org/content/369/6508/1146] for a qualitative list]). We 
 do not elaborate here on
  how to develop a set of possible
 interventions, and on how to measure their
success  on containing the
transmission rates in schools.

 \appendix

 \section{Methods}
 \subsection{Compartmental models}
In order to evaluate the effect of school opening onto
the course of the epidemic we use compartmental models,
as they proved capable of predicting the courses 
of outbreaks in many instances \cite{Brauer2017}. We start
with the simplest SIR model with unit total population, and
 two subclasses
of sizes $n_1$ and $n_2$. More features
are added later on. We make a sequence of theoretical 
claims concerning the effect of the contact rate in
the subclass representing schools. Most of the
claims are verified in suitable linear approximations of 
the SIR model; these  give very close approximations of 
non linear versions as in the entire course of the current
COVID-19
pandemic the proportion of active cases $I=I_1+I_2$ is kept
relatively low by either containment measures, lockdowns, or
 vaccinations.
Each result is then complemented with  numerical simulations.

\subsection{SIR model with two subclasses}  
We first consider the simplest model of interest, represented in terms of a coupled SIR system
\begin{equation}\label{eq:sir_model}
\begin{cases}
S_1'=-\beta_{11}S_1I_1-\beta_{12}S_1I_2\\
I_1'=\beta_{11}S_1I_1+\beta_{12}S_1I_2-I_1\\
R_1'=I_1\\
S_2'=-\beta_{21}S_2I_1-\beta_{22}S_2I_2\\
I_2'=\beta_{21}S_2I_1+\beta_{22}S_2I_2-I_2\\
R_2'=I_2
\end{cases}    
\end{equation}
 Notice that the recovery rate is the same in the two subpopulations as for COVID-19 they seem to depend on the severity of the
 infection but not directly on age \cite{Byrne}, \cite{Singanayagam2020}, \cite{Blyuss2021}, and time is rescaled so that it is equal to $1$. This makes time unit of about $10$-$14$ days \cite{Voinsky2020}. In addition, $\beta_{12}, \beta_{21}$ are generally smaller than $\beta_{11}, \beta_{22}$ \cite{PCJ17}, \cite{Kim2020}.
 We intend to compare the attack rates
 $\Delta S (\beta_{11})= S_1(t_a)-S_1(t_b)+
 S_2(t_a)-S_2(t_b)$ between two suitable times $t_a<t_b$,
 as function of the in-school transmission rate 
 $\beta_{11}$; here $\beta_{11}= 0$ corresponds
 to schools being closed, and $\beta_{11}> 0$
 corresponds to schools being open
 with varying degrees of physical distancing
 and other containment measures in place.

 \subsection{Linear approximation 
 during the initial phase of an outbreak
 or new strain upsurge}
 
A suitable linear approximation for the
initial period of the first outbreak, or of any of
the possible infection waves taking place
after a successful lockdown, is the following
\begin{eqnarray} \label{LinAppOutbrk}
\begin{cases}
S_1'=-\beta_{11}S_1(0)I_1-\beta_{12}S_1(0)I_2\label{LinOut1}\\
I_1'=(\beta_{11}S_1(0)-1)I_1+\beta_{12}S_1(0)I_2\\
S_2'=-\beta_{21}S_2(0)I_1-\beta_{22}S_2(0)I_2\\
I_2'=\beta_{21}S_2(0)I_1+(\beta_{22}S_2(0)-1)I_2\\
\end{cases}    
\end{eqnarray}
from which we extract the second and the fourth equations
for $I_1, I_2$. In 
 vector form we have
\begin{equation} \label{LinOutVect}
    \Vec{I}'=(A- \textrm{Id})\Vec{I}
\end{equation}
where $\textrm{Id}$ is the $2\times2$ identity matrix and
\begin{equation} \label{ABrelation}
    A=
    \begin{bmatrix}
    a_{11} & a_{12}\\
    a_{21} & a_{22}
    \end{bmatrix}
    :=
    \begin{bmatrix}
    \beta_{11}S_1(0) & \beta_{12}S_1(0)\\
    \beta_{21}S_2(0) & \beta_{22}S_2(0)
    \end{bmatrix}
\end{equation}
is the reproduction matrix \cite{Keeling2008}.
\begin{lemma}
\label{th:repr_H}
The solution of \eqref{LinOutVect}
is 
\begin{equation} \label{SolLinSystOutbrk}
    \Vec{I}=e^{-t}e^{At}\Vec{I}_0
    =e^{-t}(e^{\lambda_{\textrm{max}}t}\Vec{W}+e^{\lambda_{\textrm{min}}t}\Vec{V})
\end{equation}
where $\lambda_{\textrm{max}},\lambda_{\textrm{min}}$  
are the positive eigenvalues of the matrix $A$, and $\Vec{W}>0$.
\end{lemma}
The proof is in Appendix \ref{appendix_A}.

The largest eigenvalue of $A$ is
 the overall reproduction number $R_0$,\cite{Keeling2008}, 
 and the early evolution of the epidemics
 depends on the size of $\lambda_{\textrm{max}}$,
 and on the spectral gap $\lambda_{\textrm{max}}-
 \lambda_{\textrm{min}}$. This is the so-called
 slaved phase, in which the active cases
 of both populations are both lead by approximately
 the same exponential growth \cite{Keeling2008}.

 \subsection{Dependence of the largest eigenvalue of
 \(2\times2\) matrices from the first entry} 
 To get a first indication of a sudden change
 in the effect of the in-school transmission rate
 \(\beta_{11}\),
 we study the behavior of the largest
 eigenvalue of a quite general \(2\times 2\) matrix
 as function of its first entry. Let 
 \[
 \lambda'(a_{11}):= \frac{d\lambda_{max}}{d a_{11}} 
 \]
 then the following estimate holds:
 \begin{theorem} \label{EigenvaluesThm}
 Let $A=(a_{i j})$ be a $2\times2$ matrix with positive entries and let $\lambda_{max}(a_{11})>\lambda_{min}(a_{11})$ be its eigenvalues. We have that for any $\alpha\in\big(0,\frac{a_{22}}{\sqrt{a_{21}a_{12}}}\big)$
 \begin{equation}\label{eq:eigenvalues_ineq}
 \begin{aligned}
     &\frac{\lambda'_{max}(a_{22}+\alpha\sqrt{a_{12}a_{21}})-\lambda'_{max}(a_{22}-\alpha\sqrt{a_{12}a_{21}})}{\lambda'_{max}(a_{22}-\alpha\sqrt{a_{12}a_{21}})-\lambda'_{max}(0)}=:\frac{\Delta_1}{\Delta_0}\\
     &=\frac{\Delta_1}{\Delta_2}=\frac{\lambda'_{max}(a_{22}+\alpha\sqrt{a_{12}a_{21}})-\lambda'_{max}(a_{22}-\alpha\sqrt{a_{12}a_{21}})}{\lambda'_{max}(2a_{22})-\lambda'_{max}(a_{22}+\alpha\sqrt{a_{12}a_{21}})}\geq \frac{2\alpha}{\sqrt{4+\alpha^2}-\alpha}
\end{aligned}     
 \end{equation}
 \end{theorem}
 
 The proof is in Appendix  \ref{Appendix-Proof}. Figure
 \ref{PhaseTransition5} shows an example of 
 the change in derivatives.

\begin{figure}[ht]
    \centering
    \includegraphics[scale=0.7]{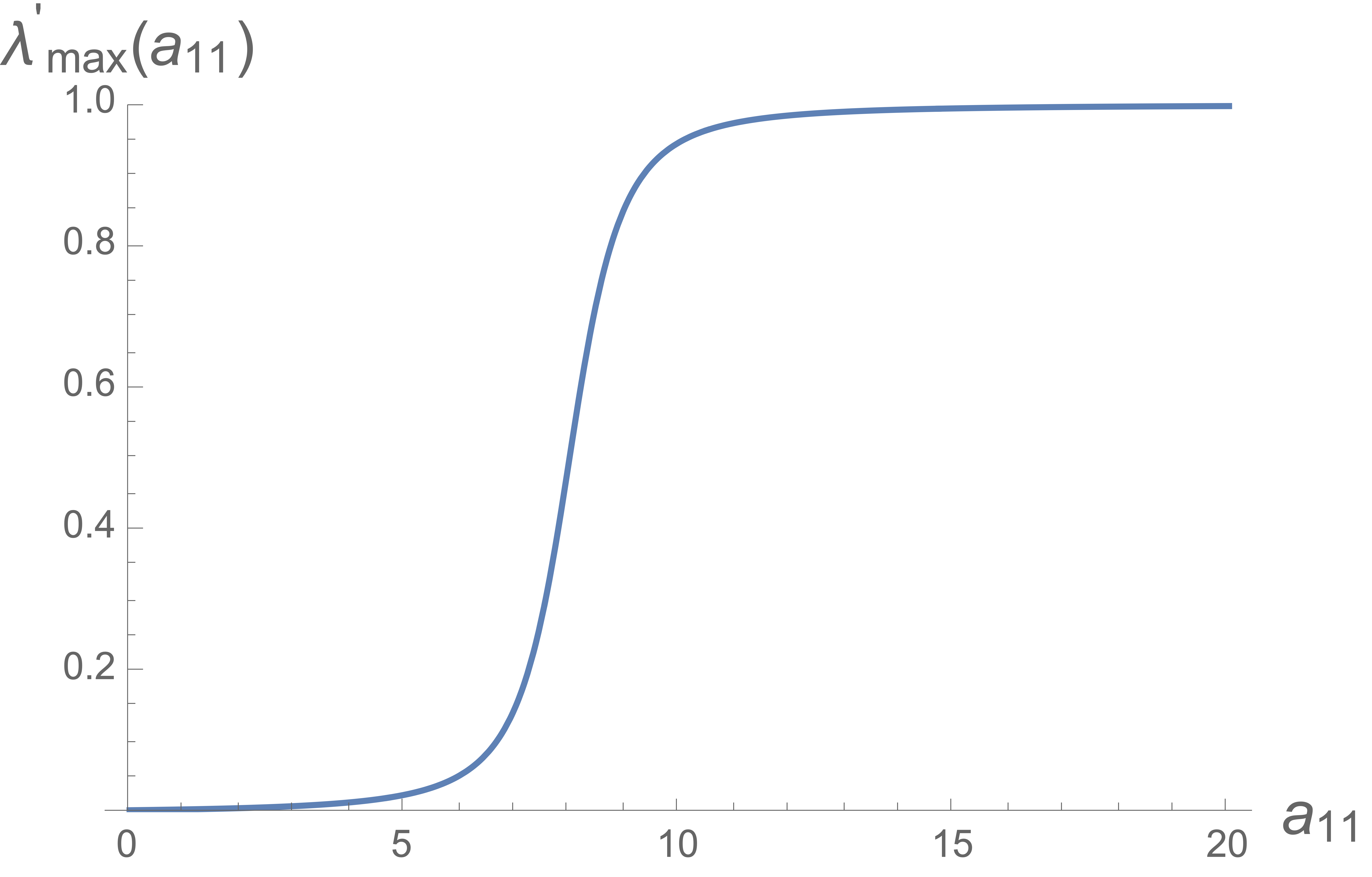}
    \caption{Derivative of the largest eigenvalue
    of $A=(a_{i j})$ with respect to $a_{11}$, with $a_{12} = a_{21} = 0.5$ and $a_{22} = 8$.
     }
     \label{PhaseTransition5}
\end{figure}
As a consequence, if, for instance, $\alpha=2$ as we will assume from now on, then
\[
    \Delta_1\geq 4.8 \Delta_0=4.8 \Delta_2.
\]
Simple calculations in Appendix \ref{Consequencies} show then that if
$ 0<a_{11}\leq a_{22}-\alpha\sqrt{a_{12}a_{21}}$,
which is realistic since  $a_{22}^2>>a_{12}a_{21}$,
\[
    \lambda_{max}(2a_{22})-\lambda_{max}(0)\geq 4.8(\lambda_{max}(a_{11})-\lambda_{max}(0)).
\]

\begin{figure}[ht]
    \centering
    \includegraphics[scale=0.6]{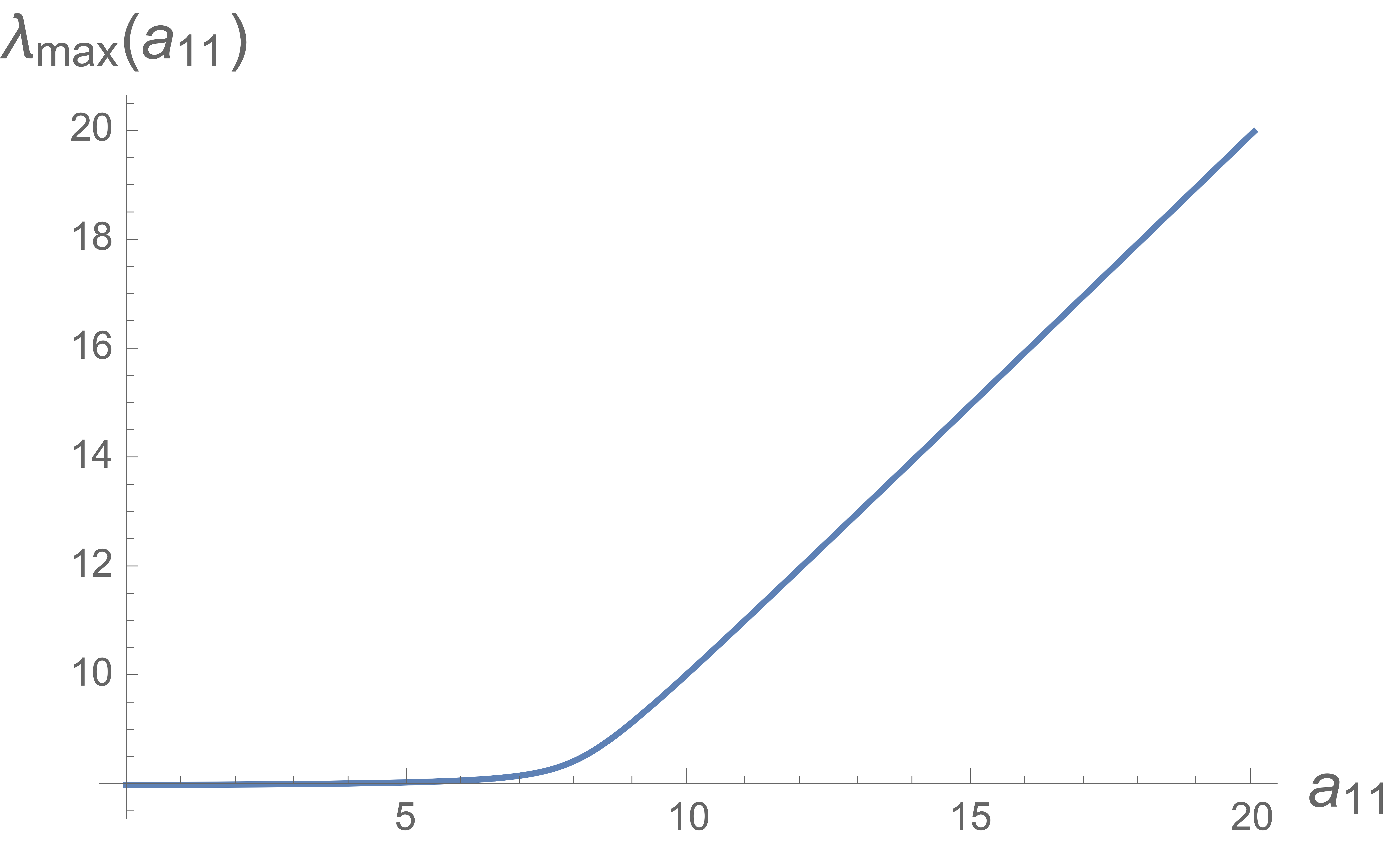}
    \caption{The largest eigenvalue
    of $A=(a_{i j})$ as function of $a_{11}$, with $a_{12} = a_{21} = 0.5$ and $a_{22} = 8$.}
     \label{PhaseTransition6}
\end{figure}

The change in the largest eigenvalue is illustrated in
Figure \ref{PhaseTransition6}.

\subsection{A phase transition for school opening during
an outbreak}

We now want to see a similar behavior in the 
active cases $I_1+I_2$ in the coupled SIR model \eqref{eq:sir_model}. 
Let 
\[
    \begin{aligned}
    \lambda_{\textrm{max}}(0)&=\lambda_0,\qquad \lambda_{\textrm{min}}(0)&=\widetilde{\lambda_0}  \\ \lambda_{\textrm{max}}(a_{22}-\alpha\sqrt{a_{12}a_{21}})&=\lambda_1,\qquad \lambda_{\textrm{min}}(a_{22}-\alpha\sqrt{a_{12}a_{21}})&=\widetilde{\lambda_1}\\
    \lambda_{\textrm{max}}(a_{22}+\alpha\sqrt{a_{12}a_{21}})&=\lambda_2,\qquad \lambda_{\textrm{min}}(a_{22}+\alpha\sqrt{a_{12}a_{21}})&=\widetilde{\lambda_2}.
    \end{aligned}
\]
and denote by $\Vec{W}^i=(W_1^i,W_2^i)$ and $\Vec{V}^i=(V_1^i,V_2^i)$, for $i=0,1,2$, the vectors $\Vec{W}$ and $\Vec{V}$ in \eqref{SolLinSystOutbrk} corresponding to $\lambda_0, \lambda_1, \lambda_2$, and $\widetilde{\lambda_0}, \widetilde{\lambda_1}, \widetilde{\lambda_2}$, respectively. 
Note that, without loss of generality, we can assume that $V^i_1>0$, for $i=0,1,2$.

Let $\Vec{I}^{\lambda_j}(t), j=0,1,2$ denote the infected at time $t$ corresponding to eigenvalues $\lambda_j,j=0,1,2$ respectively. 
\begin{corollary}\label{cor:relation_H}
For all $k>0$, there exists $T_k>0$ such that for all $t\geq T_k$
\[
    |\Vec{I}^{\lambda_2}(t)-\Vec{I}^{\lambda_0}(t)|\geq k|\Vec{I}^{\lambda_1}(t)-\Vec{I}^{\lambda_0}(t)|. 
\]
\end{corollary}
The proof is in Appendix \ref{appendix_C}. 
This identifies $a_{22}$ as the critical point
for the effect of the coefficient $a_{11}$
on the largest eigenvalue of $A$; 
by the relations in  \eqref{ABrelation},
the critical point for the effect of 
$\beta_{11}$ on the overall pandemic
is then
$\beta_{11}^*= \beta_{22} S_2(0)/S_1(0)$.

\subsection{Containment of the effect of 
school opening during an outbreak}
For $i=1,2$ and some $\overline t>0$, let 
\begin{eqnarray}
\mathcal{I}_i= \int_{0}^{\overline t}  I_i(\tau) d\tau.
\end{eqnarray}
Integrating the first and third equations of
\eqref{LinAppOutbrk}, we get
\begin{eqnarray} \label{DeltaS}
\Delta S_i&=&
S_i(0)-S_i(\overline t)
= -\int_{0}^{\overline t} S_i'(\tau) d\tau\nonumber \\
&=&\int_{0}^{\overline t} (I_i'(\tau)+ I_i(\tau))d\tau
\nonumber\\
&=& I_i(\overline t)-I_i(0)+\mathcal{I}_i.
\end{eqnarray}
On the other hand,
integrating the second and fourth equations of
\eqref{LinAppOutbrk} in $[0,\overline t]$, we get 
\begin{equation}\label{LinearOutSol}
\begin{cases}
I_{1}(\overline t)-I_1(0)=(\beta_{11}S_1(0)-1)\mathcal{I}_1+\beta_{12}S_1(0)\mathcal{I}_2\\
I_{2}(\overline t)-I_2(0)=\beta_{21}S_2(0) \mathcal{I}_1+(\beta_{22}S_2(0) -1)\mathcal{I}_2,
\end{cases}    
\end{equation}
whose solution is
\begin{equation}\label{LinearOutbrkSol}
\begin{cases}
\mathcal{I}_1=-\frac{I_1(0)-I_{1}(\overline t)  + \beta_{12} (I_2(0)-I_{2}(\overline t)) S_{1}(0) - \beta_{22} (I_1(0)-I_{1}(\overline t)) S_{2}(0) }
   {-1+\beta_{11} S_{1}(0) (1  - 
    \beta_{22} S_{2}(0))
    + \beta_{22} S_{2}(0)
    + \beta_{12} \beta_{21} S_{1}(0) S_{2}(0)}\\
 \mathcal{I}_2=-\frac{
  I_2(0)-I_{2}(\overline t) - \beta_{11}( I_2(0)-I_{2}(\overline t)) S_{1}(0) + \beta_{21} (I_1(0)-I_{1}(\overline t)) S_{2}(0) }{-1+\beta_{11} S_{1}(0) (1  - 
    \beta_{22} S_{2}(0))
    + \beta_{22} S_{2}(0)
    + \beta_{12} \beta_{21} S_{1}(0) S_{2}(0)}
\end{cases}    
\end{equation}
The attack rate  is then
\begin{eqnarray}  \label{attack}
A(\beta_{11})&:=&\frac{1}{S_1(0)+S_2(0)}(\Delta S_1+\Delta S_2)\nonumber\\
&=&\frac{1}{S_1(0)+S_2(0)}(I_{1}(\overline t)-I_1(0)+\mathcal{I}_1+I_{2}(\overline t)-I_2(0)+\mathcal{I}_2)
\nonumber\\
&=&\frac{1}{S_1(0)+S_2(0)}\Big\{\big(\beta_{11}\beta_{22}S_1(0)S_2(0)\\
&&\quad -\beta_{12}\beta_{21}S_1(0)S_2(0)\big)\big[I_1(\overline{t})-I_1(0)+I_2(\overline{t})-I_2(0)\big]\nonumber\\
&&\quad+\big(\beta_{21}S_2(0)+\beta_{11}S_1(0)\big)\big[I_1(0)-I_1(\overline{t})\big]\nonumber\\
&&\quad+\big(\beta_{12}S_1(0)+\beta_{22}S_2(0)\big)\big[I_2(0)-I_2(\overline{t})\big]\Big\}\nonumber\\
&&\quad\times \Big\{1-\beta_{11} S_{1}(0) (1  - 
    \beta_{22} S_{2}(0))
    - \beta_{22} S_{2}(0)
    - \beta_{12} \beta_{21} S_{1}(0) S_{2}(0)\Big\}^{-1}
    \nonumber
\end{eqnarray}
Notice that, to the contrary of
what happens in the next section for
the lockdown case, the value of $\beta_{11}$ 
for which the denominator is zero
does not correspond to a singularity:
this is to be expected as it differs from
$\beta_{11}^*$, and we confirmed it numerically.

Taking $\vec I$ as in \eqref{SolLinSystOutbrk}, we get 
an explicit expression for $\Delta S(\beta_{11})$.
For a fixed $\epsilon$, representing the allowed fractional
increment in the number of cases when the school is
open, the allowed bound for $\beta_{11}$ is given
by 
\begin{equation}
    A(\beta_{11})/A(0)\leq 1+\epsilon.
\end{equation}
With the parameters used in Section \ref{sub1},
this gives the mentioned value $\beta_{11}\leq 6.344$.
 
\subsection{SIR for lockdown and its linear approximation}
 
System \eqref{eq:sir_model} is suitable to model 
lockdown as well, provided that  the reproduction rate,
which is the largest 
eigenvalue of \eqref{ABrelation},
satisfies
$R_0<1$, and that  
 initial conditions taken at time 
 $\overline t$ have a more substantial number of 
cases and recovered.
A linear approximation of the 
system is possible as the overall number of active cases is
never allowed to grow beyond relatively small fractions
of the population, never more than $0.1\%$ in most
countries.

With these conditions, a linear approximation is
\begin{equation}\label{LinearLockDown}
\begin{cases}
I_{1}'=\overline\beta_{11}S_1(\overline t)I_{1}+\overline\beta_{12}S_1(\overline t)I_{2}-I_{1}\\
I_{2}'=\overline\beta_{21}S_2(\overline t) I_{1}+\overline\beta_{22}S_2(\overline t) I_{2}-I_{2}\\
I_{1}(\overline t), I_{2}(\overline t) >0,
\end{cases}    
\end{equation}
 with $S_1(\overline t)+S_2(\overline t)
 +I_{1}(\overline t)+I_{2}(\overline t)<1$.
 
\begin{figure}[ht]
    \centering
    \includegraphics[scale=0.5]{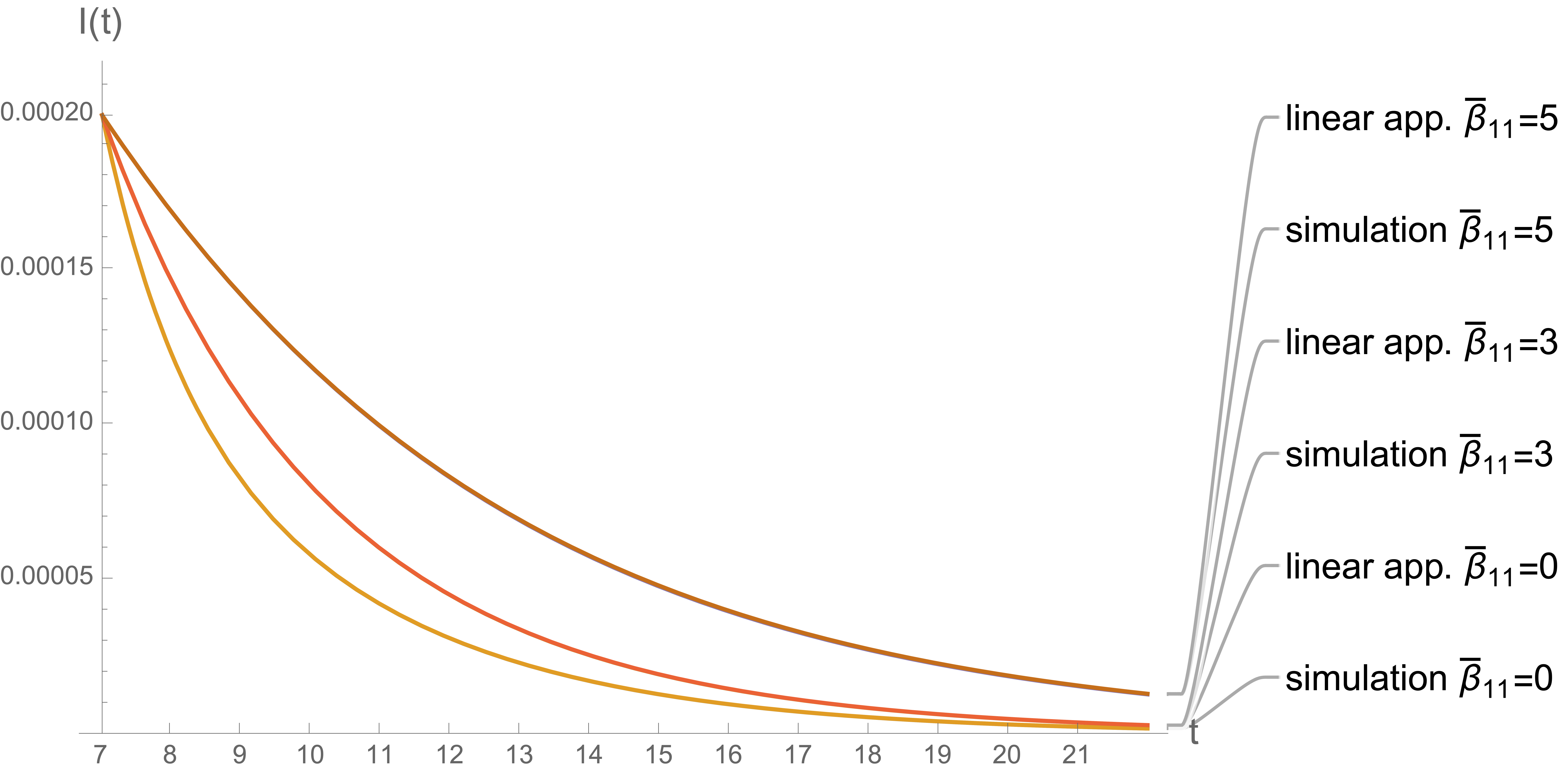} 
    \caption{Effectiveness of the 
linear approximation of the SIR model for lockdown;
the figure shows the total active cases numerically 
simulated with realistic parameters and varying $\overline\beta_{11}$:
in each test, simulations from the differential system and from
its linear approximation are indistinguishable.
     }
     \label{PhaseTransition8}
\end{figure}

 Figure \ref{PhaseTransition8} illustrates via a 
 simulation for $S_1(\overline t)=0.15<0.2,
 S_2(\overline t)=0.7<0.8,
 I_{1}(\overline t)=I_{2}(\overline t)=10^{-4},
  \overline\beta_{12}= \overline\beta_{21} =0.25, \overline\beta_{22}=1 $,
  and $\overline\beta_{11}=0,3,5$, the closeness of the linear 
  approximation. The total number of cases simulated from the differential system
and the linear approximation are not distinguishable
in the figure for all values of $\overline\beta_{11}$; the same holds for each
subpopulation.

\subsection{Allowed level of school transmission 
for a successful lockdown}
Let us assume that a lockdown is applied
from time $\overline t$ to $ \tilde t$
that successfully eradicate the virus; 
hence with $I_i(\tilde t) \approx 0$
for $i=1,2$.
From the mathematical point of 
view, we can take the eradication time to 
be $+\infty$ as the dynamical system reaches an equilibrium 
 with no active cases and does not change afterwards.
 Hence we consider $\Delta S_i=S_i(\infty)-S_i(\tilde t)
 \approx S_i(\overline t)-S_i(\tilde t)$.
 Formulas  \eqref{DeltaS} -\eqref{attack} apply, with
 $0$ and $\overline t$ replaced by $\overline t$ and
 $\infty$, respectively; the quantities
 ${\mathcal{I}}_i, \beta_{i,j}$, for $i,j=1,2$,
 decorated by an overscore; and 
 $I_1(\infty)=I_2(\infty)=0$.

The denominator of \eqref{attack} with the above
changes is singular for 
\begin{eqnarray} \label{BetaCritLock}
\overline{\beta}_{11}^{*}=\frac{1}{S_{1}(\overline t)}
-\frac{\overline{\beta}_{12} \overline{\beta}_{21} S_{2}(\overline t)}{  (1 - \overline{\beta}_{22} S_{2}(\overline t))}.
\end{eqnarray}
 The numerator at $\overline{\beta}_{11}=\overline{\beta}_{11}^{*}$,
 on the other hand, is not identically zero; this is  seen by substituting 
 the value  \eqref{BetaCritLock} for $\overline{\beta}_{11}$
 in \eqref{attack}, with the adaptations listed above:
 after some algebra, carried out in Mathematica\textsuperscript{TM}, the 
 numerator is seen to 
 equal
 \begin{eqnarray}
 \frac{(1 + \overline{\beta}_{21} S_{2}(\overline t))
 ( \overline{\beta}_{12} I_{2}(\overline t)
   S_{1}(\overline t)+I_1(\overline t)(1 - \overline{\beta}_{22} S_{2}(\overline t)) )}{(1 - \overline{\beta}_{22} S_{2}(\overline t))} \approx -1.6\times 10^{-5};
 \end{eqnarray}
 the numerical value is computed with the values indicated in Section \ref{PhaseTransLock},
namely that 
\begin{eqnarray} \label{ValuesLock1}
\overline\beta_{12}= \overline\beta_{21} =0.25, \overline\beta_{22}=1, \epsilon=0.3 
\end{eqnarray}
and choosing as initial condition at 
 $t=\bar t=5$ the total number of susceptible and infected obtained from the outbreak scenario, that is
\begin{eqnarray} \label{ValuesLock2}
S_1(\overline t)\approx0.1996, S_2(\overline t)
\approx0.7981,I_1(\overline t)\approx1.601\times 10^{-4},  I_2(\overline t)
\approx7.9555\times 10^{-4}.
\end{eqnarray}
This indicates that the value in \eqref{BetaCritLock}
is where the linear approximation breaks down,
indicating a transition of phases.

\medskip
 
In addition, if we require that school opening does not 
affect more than a certain percentage the
overall incidence proportion by asking that
\[
A(\overline{\beta}_{11})\leq(1+\epsilon)
A(0)
\]
for some $\epsilon>0$, then after some algebra,
carried out in Mathematica\textsuperscript{TM},we get
that 
\begin{equation} \label{CapitalF}
\begin{aligned}
\overline{\beta}_{11}\leq F(\epsilon)&=
\Big[\epsilon \Big(-1 + \overline{\beta}_{22} S_2(\overline{t}) +
     \overline{\beta}_{12} \overline{\beta}_{21} S_1(\overline{t}) S_2(\overline{t})\Big) \Big(\overline{\beta}_{12}I_2(\overline{t}) S_1(\overline{t}) \\
     &+ \overline{\beta}_{21} I_1(\overline{t}) S_2(\overline{t}) + \overline{\beta}_{22} I_2(\overline{t}) S_2(\overline{t}) +
     \overline{\beta}_{12} \overline{\beta}_{21} I_1(\overline{t}) S_1(\overline{t}) S_2(\overline{t}) \\
     &+
     \overline{\beta}_{12} \overline{\beta}_{21} I_2(\overline{t}) S_1(\overline{t}) S_2(\overline{t})\Big)\Big]\times \Big\{S_1(\overline{t})\Big[-I_1(\overline{t})-
     \overline{\beta}_{12}I_2(\overline{t}) S_1(\overline{t})\\
     &- \epsilon\overline{\beta}_{12} I_2(\overline{t})S_1(\overline{t}) -
     \overline{\beta}_{21}I_1(\overline{t})S_2(\overline{t})+ 2 \overline{\beta}_{22} I_1(\overline{t})S_2(\overline{t}) -\epsilon
     \overline{\beta}_{21} I_1(\overline{t})S_2(\overline{t})\\
     &- \epsilon\overline{\beta}_{22} I_2(\overline{t})S_2(\overline{t}) -
     \epsilon\overline{\beta}_{12} \overline{\beta}_{21}I_1(\overline{t}) S_1(\overline{t}) S_2(\overline{t})-
     \overline{\beta}_{12} \overline{\beta}_{21} I_2(\overline{t}) S_1(\overline{t}) S_2(\overline{t})\\
     &+
     \overline{\beta}_{12} \overline{\beta}_{22} I_2(\overline{t})S_1(\overline{t}) S_2(\overline{t})-\epsilon
     \overline{\beta}_{12} \overline{\beta}_{21} I_2(\overline{t}) S_1(\overline{t}) S_2(\overline{t}) \\
     &+\epsilon
     \overline{\beta}_{12} \overline{\beta}_{22} I_2(\overline{t})S_1(\overline{t}) S_2(\overline{t}) +
     \overline{\beta}_{21} \overline{\beta}_{22}I_1(\overline{t}) S_2^2(\overline{t}) - \overline{\beta}^2_{22}I_1(\overline{t}) S^2_2(\overline{t}) \\
     &+
     \epsilon \overline{\beta}_{21} \overline{\beta}_{22} I_1(\overline{t}) S^2_2(\overline{t}) + \epsilon \overline{\beta}^2_{22} I_2(\overline{t}) S^2_2(\overline{t}) \\
     &+\epsilon
     \overline{\beta}_{12} \overline{\beta}_{21} \overline{\beta}_{22} I_1(\overline{t})S_1(\overline{t})S^2_2(\overline{t}) +
     \epsilon \overline{\beta}_{12} \overline{\beta}_{21} \overline{\beta}_{22} I_2(\overline{t}) S_1(\overline{t}) S^2_2(\overline{t})\Big]\Big\}^{-1}
\end{aligned}
\end{equation}
With the values as in \eqref{ValuesLock1} and \eqref{ValuesLock2}
we get 
$F(0.3) \approx 2.9944$.
 \medskip

We compare this expression with 
the bound in \eqref{FirstCriticalPoint}, in  some
numerical examples.

\subsection{A complete outbreak-lockdown cycle }

A confirmation of  the behavior of the effect of 
 school opening on one outbreak-lockdown cycle
 is shown here via a direct simulation.

Continuing the  numerical example of 
Section \ref{PhaseTransLock}, suppose a lockdown 
is imposed starting from
$\overline t=5$, and a $50\%$ reduction is achieved in the 
transmission rates different from $\beta_{11}$; Figure \ref{Recap},
$t \in [5,18]$, shows that as the outbreak is resolved
after the lockdown, the cumulative number of cases is close to that at $\overline \beta_{11}=0 $
when  $\beta_{11}$ and $\overline \beta_{11} $ are 
close enough to the origin, and sharply deviates 
otherwise.

\begin{figure}[ht]
    \centering
    \includegraphics[scale=0.5]{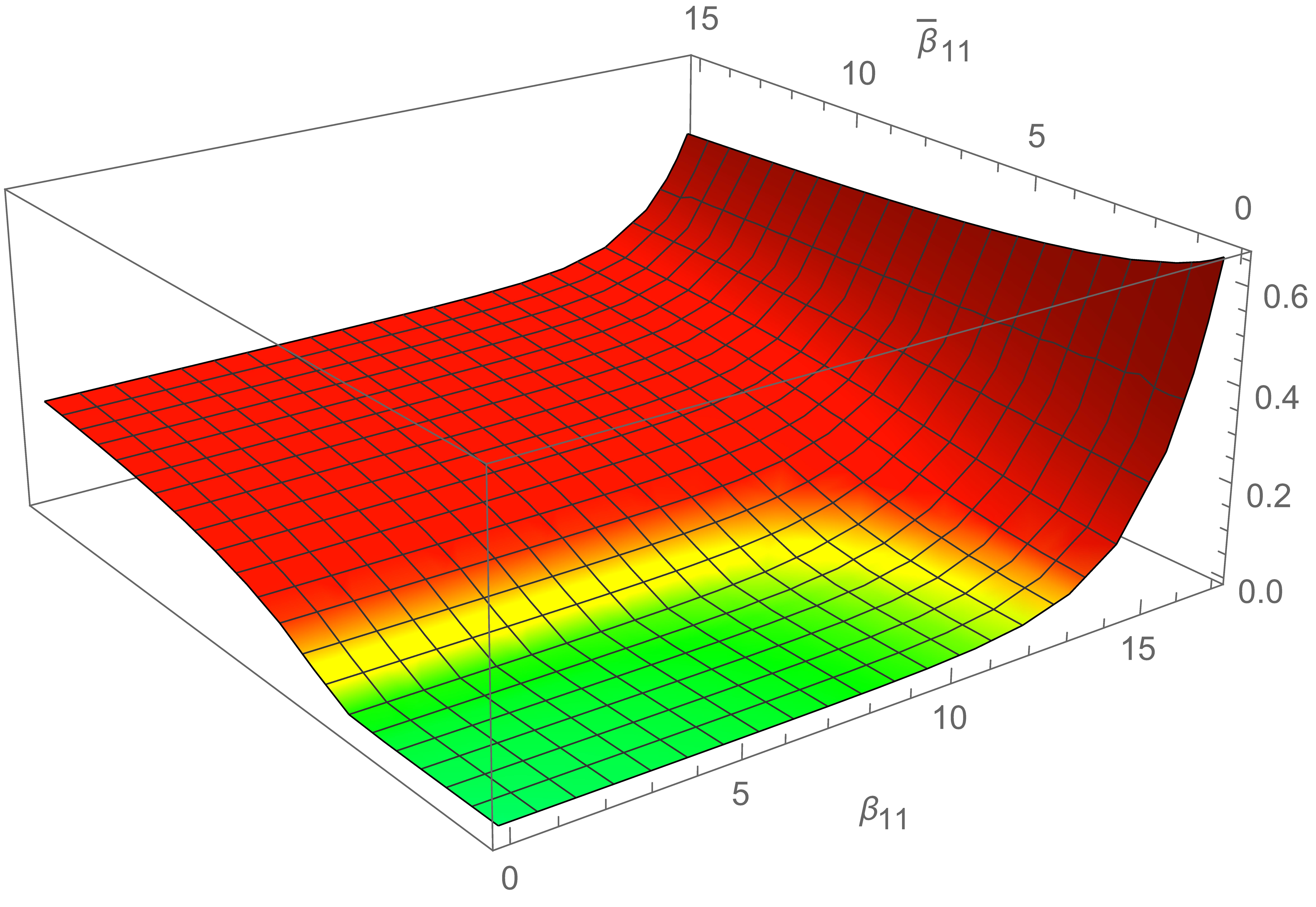}
    \caption{Total number of cases at resolution of outbreak for varying
    values of $\beta_{11}$ and $\overline \beta_{11} $.
     }
     \label{PhaseTransition3}
\end{figure}

\subsection{SIR with two subpopulations and vaccination}
Vaccination is included in the model assuming
that the adult population is vaccinated at 
a constant rate $v$:
\begin{equation}\label{eq:sir_model_vacc}
\begin{cases}
S_1'=-\widetilde{\beta}_{11}S_1I_1-\widetilde{\beta}_{12}S_1I_2\\
I_1'=\widetilde{\beta}_{11}S_1I_1+\widetilde{\beta}_{12}S_1I_2-I_1\\
R_1'=I_1\\
S_2'=-\widetilde{\beta}_{21}S_2I_1-\widetilde{\beta}_{22}S_2I_2-v S_2\\
I_2'=\widetilde{\beta}_{21}S_2I_1+\widetilde{\beta}_{22}S_2I_2-I_2\\
R_2'=I_2+v S_2.
\end{cases}    
\end{equation}
see, e.g. \cite{Libotte et al.20},\cite{De la Sen et. al}.
Here, for seek of simplicity we assume that the time $\tilde t$ at which the vaccination
program begins corresponds to $t=0$.

We are not aware of a way to 
explicitly express the  solutions of this system.

\subsection{A linear approximation to SIR with vaccination}
\begin{figure}[ht]
    \centering
    \includegraphics[scale=0.5]{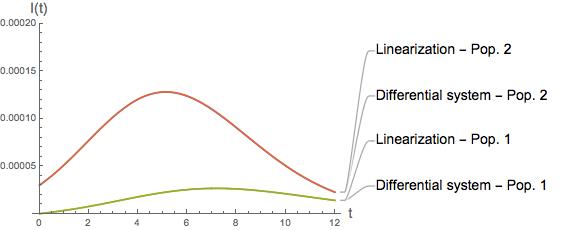}
    \caption{Effectiveness of the 
linear approximation of the SIR model with vaccination;
the figure shows the total active cases numerically 
simulated with realistic parameters,
$\widetilde{\beta}_{11}=3,\widetilde{\beta}_{22}=2$:
simulations from the differential system and
from its linear approximation are indistinguishable.
     }
     \label{PhaseTransition7}
\end{figure}

In order to analyze \eqref{eq:sir_model_vacc}
we develop a linearization. Notice that in the linear
approximation for the initial phase of an SIR model,
the terms
$\widetilde{\beta}_{i1}S_iI_1+\widetilde{\beta}_{i2}S_iI_2$
are taken to be zero for both $i=1,2$. With the 
same assumption in the vaccination case, we get
the equation 
$S_2'=-v S_2$: we therefore use the solution to 
this equation as linear approximation of $S_2(t)$.
This leads to  the following linearization
\begin{equation}\label{eq:sir_model_vacc_linear}
\begin{cases}
S_1'=-\widetilde{\beta}_{11}S_1(0) I_1-\widetilde{\beta}_{12}S_1(0)I_2\\
I_1'=\widetilde{\beta}_{11}S_1(0)I_1+\widetilde{\beta}_{12}S_1(0)I_2-I_1\\
R_1'=I_1\\
S_2'=-\widetilde{\beta}_{21}S_2(0) e^{-v t}I_1-\widetilde{\beta}_{22}S_2(0) e^{-v t}I_2-v S_2\\
I_2'=\widetilde{\beta}_{21}S_2(0) e^{-v t}I_1+\widetilde{\beta}_{22}S_2(0) e^{-v t}I_2-I_2\\
R_2'=I_2+v  S_2.
\end{cases}    
\end{equation}
 From \eqref{eq:sir_model_vacc_linear} we extract
\begin{equation}\label{eq:sir_model_vacc_linear extract}
\begin{cases}
I_1'=\widetilde{\beta}_{11}S_1(0)I_1+\widetilde{\beta}_{12}S_1(0)I_2-I_1\\
I_2'=\widetilde{\beta}_{21}S_2(0) e^{-v t}I_1+\widetilde{\beta}_{22}S_2(0) e^{-v t}I_2-I_2.
\end{cases}    
\end{equation}
Figure \ref{PhaseTransition7} shows one instance 
of the effectiveness of the 
linear approximation with realistic parameters.

\vskip 2cm

\subsection{Evidence of a phase transition in $\widetilde{\beta}_{11}$ with 
critical point $1/{S_1(0)}$ during vaccination}
We proceed by using the linear approximation to evaluate 
  the attack rates as function of  
$\widetilde{\beta}_{11} $; in particular, we
focus on the one for the external population $2$.
Our calculation is done
 recursively, as shown in Appendix \ref{appendix_E}.
The following theorem summarizes the calculation.
\begin{theorem} \label{attack-rates}
Assume that $I_1,I_2$ are integrable in $[0,+\infty)$
and let
\begin{eqnarray}
\widetilde{\mathcal{I}}_1&=&\widetilde{\mathcal{I}}_1(\widetilde{\beta}_{11})=\int_0^{+\infty}I_1(t) dt, \quad
\widetilde{\mathcal{I}}_2=\widetilde{\mathcal{I}}_2(\widetilde{\beta}_{11})=\int_0^{+\infty}I_2(t) dt,\\
p_{k+1}&=&I_2(0)+
\frac{\widetilde{\beta}_{21}S_{2}(0) I_1(0)} {(k+1)v-\widetilde{\beta}_{11}S_1(0)+1}
\\
q_{k+1}&=&
\widetilde{\beta}_{22} S_{2}(0)+
\frac{\widetilde{\beta}_{12}\widetilde{\beta}_{21} S_{1}(0) S_{2}(0))}{(k+1)v-\widetilde{\beta}_{11}S_1(0)+1}
\end{eqnarray}
for $k=0, 1, \dots$.
We have that
\begin{eqnarray}
\widetilde{\mathcal{I}}_1&=&\widetilde{\mathcal{I}}_1(\widetilde{\beta}_{11})= \frac{\widetilde{\beta}_{12}S_{1}(0) \widetilde{\mathcal{I}}_2+I_1(0)}{1-\widetilde{\beta}_{11}S_1(0)}
\label{Jvalue} \\ 
\widetilde{\mathcal{I}}_2&=&\widetilde{\mathcal{I}}_2(\widetilde{\beta}_{11})= \sum_{i=1}^{\infty} p_i\Bigg(\prod_{r=1}^{i-1}
\frac{q_r}{r v+1}\Bigg)
\label{Hvalue} 
\end{eqnarray}

\end{theorem}
The proof is in Appendix \ref{appendix_E}, where we also give an 
explicit expression for $\widetilde{\mathcal{I}}_2$ in terms of
hypergeometric functions.

To compute the attack rate for the vaccination case
observe that the change in active cases is given
by $I_i'$, and the change in recovered cases is $I_i$,
$i=1, 2$;  
the change in infected is then
$I'+I$, and the attack rate is
\begin{eqnarray}
A(\widetilde{\beta}_{11}) &=&
\frac{1}{S_1(0)+S_2(0))}\int_0^\infty (I'_1+I_1+I'_2+I_2)dt
\\ &=&
\frac{1}{S_1(0)+S_2(0))}(-I_1(0)+\widetilde{\mathcal{I}}_1-I_2(0)+\widetilde{\mathcal{I}}_2) \nonumber \label{DeltaS2}
\end{eqnarray}
The attack rate $A$ is divergent as
$\widetilde{\beta}_{11}$ approaches $\widetilde{\beta}_{11}^{*}=
1/{S_1(0)}$, see \eqref{Jvalue}, which 
is an indication that the linear approximation
breaks down, and that this value is likely to be the critical
point. 

In order to contain the increase in the total cases by 
no more than a proportion $\epsilon$
we need $\widetilde{\beta}_{11}$ satisfying
\begin{equation} \label{UpperBoundVacc}
    A (\widetilde{\beta}_{11})\leq A(0) (1+\epsilon).
\end{equation}

\subsection{Estimate of the
peak time during vaccination}
A further evidence of the critical point
is obtained by an estimate of the
peak time of the infection from
\eqref{eq:sir_model_vacc_linear extract}.
Assuming that the active cases in the two subpopulations
peak at approximately the same time  $\overline t$,
we set $I_1'(t)=I_2'(t)=0$.
The solution is 
\begin{equation}
    \overline t= 
    \frac 1 v \log{
   \frac{-\widetilde{\beta}_{12}S_1(0)\widetilde{\beta}_{21}S_2(0)+
   \widetilde{\beta}_{22}S_2(0)(1-\widetilde{\beta}_{11})S_1(0)}{1-\widetilde{\beta}_{11}S_1(0)} }.
\end{equation}
Hence, the peak time also diverges at 
$\widetilde{\beta}_{11}=\widetilde{\beta}_{11}^{*}$.

\subsection{Simulations of the phase 
transition during vaccination}

With the realistic values of the parameters
used previously,
 \[S_1(0)=0.198872,
 S_2(0)=0.794451,
 I_{1}(0)=1.97281 \times 10^{-5},I_{2}(0)=9.55298 \times 10^{-5}\]
 and
\[ \beta_{12}= \widetilde{\beta}_{21} =0.5, \widetilde{\beta}_{22}=2,
  v=0.1,\]
 we have 
 \[\widetilde{\beta}_{11}^{*}=1/{S_1(0)}=5.02836.\]
 
Figure \ref{Recap} for $t > 18$ illustrates
a simulation of the differential system, where it is seen
that $\widetilde{\beta}_{11}^{*}=5.02836$
is the critical point for the influence of 
school opening on the overall epidemic.

To achieve a sensible containment take
 $\epsilon=0.3$, in which case \eqref{UpperBoundVacc}
 gives $\widetilde{\beta}_{11}<3.03111$, a bound also visible
 in Figure \ref{Recap} for $t>18$.

 \subsection{A SPIAR model} \label{SPIAR}
 To illustrate how a phase transition mechanism
 also appears in more  elaborate and realistic models, 
 we develop and simulate one example.
 
 We introduce the compartments of susceptibles $S_i$ (not subjected to any virus transmission), presymptomatic $P_i$ (infected in incubation period), asymptomatic $A_i$ (infected not showing symptoms after incubation), infected $I_i$ and recovered $R_i$ for $i=1,2$ corresponding to the two subpopulations. A 
 corresponding system could read as follows:
 
 \begin{eqnarray} 
\frac{dS_{1}}{dt} & =&  
 -  S_1 \left(\beta_{11} (P_1+s A_1)+\beta_{12}(P_2+A_2+\xi I_2)\right)
\nonumber \\
\frac{dP_{1}}{dt} & =&  S_1 \left(\beta_{11} (P_1+s A_1)+\beta_{12}(P_2+A_2+\xi I_2)\right)-\kappa P_1  
\nonumber \\
\frac{dI_{1}}{dt} & =& \epsilon_1\kappa P_1- I_1 
\nonumber \\
\frac{dA_{1}}{dt} & =& (1-\epsilon_1)\kappa P_1- A_1
\nonumber \\
\frac{dR_{1}}{dt} & =& (I_1+A_1)
\nonumber \\
\frac{dS_{2}}{dt} & =&  
 -  S_2 \left(\beta_{21}(P_1+A_1+\xi I_1)+\beta_{22}(I_2+P_2+A_2) -v\right) 
\nonumber \\
\frac{dP_{2}}{dt} & =& S_2 \left(\beta_{21}(P_1+A_1+\xi I_1)+\beta_{22}(I_2+P_2+A_2)\right) -\kappa P_2  
\nonumber \\
\frac{dI_{2}}{dt} & =& \epsilon_2\kappa P_2- I_2 
\nonumber \\
\frac{dA_{2}}{dt} & =& (1-\epsilon_2)\kappa P_2- A_2
\nonumber \\
\frac{dR_{2}}{dt} & =& (I_2+A_2) +v S_2
\nonumber
\end{eqnarray}
where the parameters $s$, $\xi$ represent  the fractions of asymptomatic encountered at school and of undetected infected
individuals, respectively; $\epsilon_1$, $\epsilon_2$ are
the fractions of symptomatic in subclass $1$ and subclass $2$, respectively; $\kappa$ the rate of exit from latency period.
The recovery rate $\gamma$ 
is normalized to $1$ as before, and $v=0$ if there is no
ongoing vaccination.

Parameters have been calibrated as
given in Table \ref{ModelParameters}
following standard
estimations appearing in literature and 
data studies \cite{Oran2020,Davies2020,Dong2020}.  
\begin{table}[htbp]
  \centering
  \caption{Recap of the model parameters and their 
selected values for each scenario in SPIAR model.}
    \begin{tabular}{lrccc}
         \multicolumn{1}{l}{Parameter} & \multicolumn{1}{l}{Outbreak} & \multicolumn{1}{l}{Lockdown} & \multicolumn{1}{l}{Vaccination}
         \\ \hline \vspace{0.1cm}
    $\beta_{22}$ & $2 $ & $0.5 $ & $1.5$  \\ \vspace{0.1cm}
    $\beta_{12}=\beta_{21}$ & $0.25 $ & $0.1 $ & $0.25$  \\ \vspace{0.1cm}
    $s$ & $0.9$ & $0.9$ & $0.9$   \\ \vspace{0.1cm}
    $\xi$ & $0.3 $ & $0.3 $ & $0.3 $    \\ \vspace{0.1cm}
    $k$ & $1$ & $1$ & $1$   \\ \vspace{0.1cm}
    $\epsilon_1$ & $0.1 $ & $0.1 $ & $0.1 $   \\ \vspace{0.1cm}
    $\epsilon_2$ & $0.3 $ & $0.3 $ & $0.3 $  \\ \vspace{0.1cm}
    $v$ & $0 $ & $0 $ & $0.2 $    \\ \hline
    \end{tabular}%
  \label{ModelParameters}%
\end{table}%

\begin{figure}
    \centering
    \includegraphics[scale=0.35]{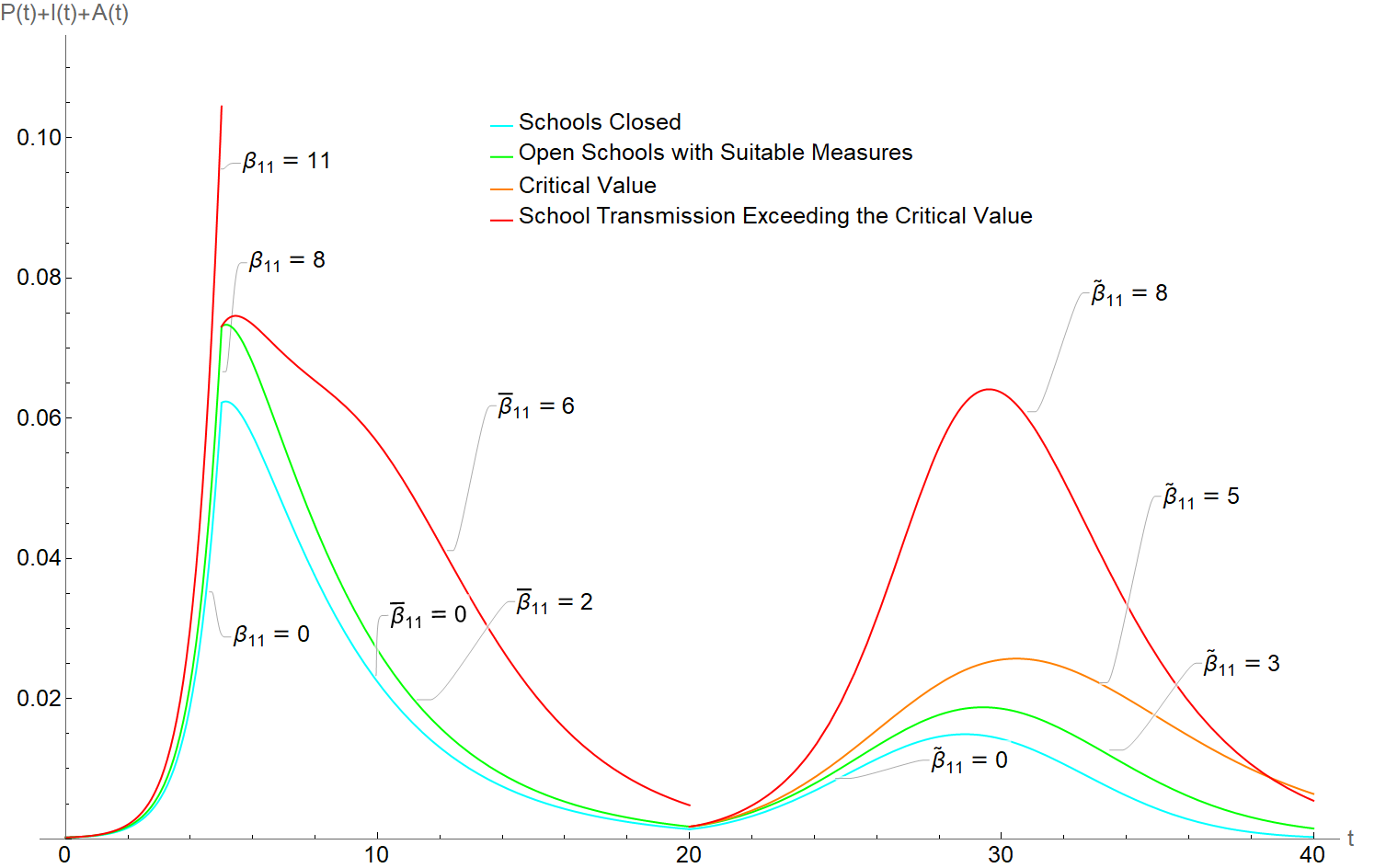}
    \caption{Daily cases for the three scenarios 
    of outbreak, lockdown, and vaccination in an SPIAR model. In each case, there is critical values for the in-school transmission rate $\beta_{11}$, as for SIR model. Cyan curve is with closed schools, green for safe opening, orange for critical values, red for values above criticality.}
    \label{fig:SPIAR}
\end{figure}

Figure \ref{fig:SPIAR} shows how  phase transitions appear also in the SPIAR model. Here, the active cases are given by the sum $P(t)+I(t)+A(t):=P_1(t)+I_1(t)+A_1(t)+P_2(t)+I_2(t)+A_2(t)$,
and  three scenarios are considered, as before: outbreak, lockdown, and vaccination. In Figure \ref{fig:SPIAR}, the cyan curve corresponds to 
closed schools, while the green one is a subcritical 
pattern; the red curves, instead, show the risk that
the pandemic
spirals out of control because of insufficiently
controlled school opening.

 \vskip 1cm

\subsection{Other case studies} \label{CaseStudies}

A survey of many detailed, data driven studies
related to the effect of school opening during the
pandemic shows traces of phases transition in all  of them.

\cite{Espana2021} uses a detailed
compartmental model
calibrated on 
mortality and other estimated and observed data
in Bogot\`a, Colombia,
during the whole 2020. The study develops various 
scenarios of school reopening, and evaluates its impact;
the phase transition described in our work
appears clearly in Figures $4$ and $5$ of \cite{Espana2021}: one
can see that
up to $50\%$ capacity the effect of opening
schools is almost negligible, while it
becomes substantial above $75\%$ capacity.
This leads to the conclusion that there has to be
a critical point between these values.

The appearance of a phase transition phenomenon in
\cite{Rozhnova et. al} has been discussed at
length in Section
\ref{EvidenceData}.

\cite{Yuan2021} uses a detailed compartmental model and data from
the second semester 2020 in Toronto, 
an outbreak context, to estimate the 
likely impact of school opening,
also discussed in Section
\ref{EvidenceData}.

\cite{Domenico2021}
analyses French data for late Spring 2020 in order to 
predict the effect of various forms of 
school reopening after the end
of the lockdown;  the paper only makes
predictions for short periods (see Figures $3$ and $5$),
and the exact details of the contacts and transmission
rates, which are partially estimated and partly obtained
from previous measurements, are not provided;
still, one can see, especially in 
Figure $5$, that the transmission rates 
are supercritical, and school opening determines
a sharp increase in the overall epidemic spreading.

Among the statistical papers,  \cite{Iwata2020}
performs a
Time-series analyses  using the Bayesian method
on Japanese data collected  during the initial lockdown,
and suggests that school closure 
did not appear to decrease the incidence of COVID-19.
\cite{Matzinger2020}, on the other hand,
uses US data from the early stages of the
outbreak, and regression analysis;
the study finds empirical evidence suggesting that 
school closings dropped infection rate to half:
we can interpret this as a sign that 
transmission rate in schools 
was supercritical at that time.

 An 
analysis of data gathered by a surveillance of 
COVID-19 cases in students and staff after reopening 
of schools across England 
showed that 
in-school infections were much less influential
than external ones \cite{Ismail2020};
a study of Italian data
from early Fall 2020, a period of
 low epidemic incidence, also showed very little
 transmission taking place in schools
 \cite{Larosa2020}. These were  
 typical 
examples of subcritical in-school transmission rate,
probably due to the segmented or very controlled reopening of
schools.

\section{Proof of Lemma \ref{th:repr_H}}
\label{appendix_A}

\begin{proof}

Consider, first, 
\begin{equation}\label{def: H}
\Vec{H}=e^t \Vec{I}. 
\end{equation}
Then
\[
    \frac{d \Vec{H}}{dt}= e^{ t}\Vec{I}+e^{ t}(A-\textrm{Id}) \Vec{I}=e^{ t}A\Vec{I}=A e^{ t}\Vec{I}=A \Vec{H}
\]
with $ \Vec{H}(0)=\Vec{I}(0):=\Vec{I}_0$. Since
\[
    \Vec{H}=e^{A t} \Vec{H}_0=e^{At}\Vec{I}_0
\]
we get that 
\begin{equation}\label{eq:relation_H_I}
    \Vec{I}=e^{-t}e^{At}\Vec{I}_0.
\end{equation}

We, now, show that
\begin{equation}\label{eq:expr H}
    \Vec{H}(t)=e^{\lambda_{\textrm{max}}t}\Vec{W}+e^{\lambda_{\textrm{min}}t}\Vec{V}
\end{equation}
where $\Vec{W}>0$. Then, the assertion of the theorem, which is related to $\Vec{I}$, follows immediately from \eqref{eq:relation_H_I}.

The rate of growth in the initial exponential phase depends on the largest eigenvalue of the reproduction matrix \eqref{ABrelation}.
The result is an immediate consequence of the Perron Frobenius theorem. In fact, since all the elements of $A$ are positive, then the eigenvector $\Vec{\xi}$ associated to $\lambda_{\textrm{max}}$ has positive components while the eigenvector $\Vec{\eta}$ associated to $\lambda_{\textrm{min}}$ has at least one negative component. Since
\[
    \Vec{H}(t)=e^{\lambda_{\textrm{max}}t}
    \begin{bmatrix}
    \frac{1}{\alpha}(\eta_2\xi_1 I_{1}(0)-\eta_1\xi_1 I_{2}(0))\\
    \frac{1}{\alpha}(\eta_2\xi_2 I_{1}(0)-\eta_1\xi_2 I_{2}(0))\\
    \end{bmatrix}
    +
    e^{\lambda_{\textrm{min}}t}
    \begin{bmatrix}
    \frac{1}{\alpha}(-\eta_1\xi_2 I_{1}(0)+\eta_1\xi_1 I_{2}(0))\\
    \frac{1}{\alpha}(-\eta_2\xi_2 I_{1}(0)+\eta_2\xi_1 I_{2}(0))\\
    \end{bmatrix}
\]
where 
\[
    \alpha=\textrm{det}
    \begin{bmatrix}
    \xi_1 & \eta_1 \\
    \xi_2 & \eta_2
    \end{bmatrix}
    =\xi_1\eta_2-\xi_2\eta_1.
\]
Then, if, for example, $\eta_1<0$ we have $\eta_2\geq 0$ hence $\alpha>0$, that is
\[
\begin{aligned}
    W_1&=\frac{1}{\alpha}(\eta_2\xi_1I_{1,0}-\eta_1\xi_1I_{2,0})>0\\
    W_2&=\frac{1}{\alpha}(\eta_2\xi_2 I_{1,0}-\eta_1\xi_2I_{2,0})>0.\\
\end{aligned}    
\]
Analogously if $\eta_1\geq 0$, then $\eta_2<0$, hence $\alpha<0$ and again $W_1>0$ and $W_2>0$.
\end{proof}

 \section {Proof of Theorem \ref{EigenvaluesThm}}
 \label{Appendix-Proof}
 \begin{proof}
 Note first that 
 \[
 \begin{aligned}
     \lambda_{max,min}=\frac{\text{tr}(A)\pm \sqrt{(\text{tr}(A))^2-4\text{det}(A)}}{2}&=\frac{a_{11}+a_{22}\pm \sqrt{(a_{11}-a_{22})^2+4a_{12}a_{21}}}{2}\\
     &> \frac{a_{11}+a_{22}\pm (a_{11}-a_{22})}{2}> 0,
\end{aligned}     
\]
 since all the entries of the matrix $A$ are positive.
Therefore $\lambda_{max},\lambda_{min}>0$. Furthermore
\[
  \lambda'(a_{11})=\frac{1}{2}\Bigg( 1+\frac{a_{11}-a_{22}}{\sqrt{(a_{11}-a_{22})^2+4a_{21}a_{12}}}\Bigg)>0
\]
and
\[
  \lambda''(a_{11})=\frac{2 a_{12}a_{21}}{((a_{11}-a_{22})^2+4a_{21}a_{12})^{3/2}}>0.
\]
We have
\[
    \lambda'_{max}(0)=\frac{1}{2}\Bigg(1-\frac{a_{22}}{\sqrt{a_{22}^2+4a_{12}a_{21}}} \Bigg)
\]
and
\[
\begin{aligned}
    \lambda'_{max}(a_{22}-\alpha\sqrt{a_{12}a_{21}})&=\frac{1}{2}\Bigg(1-\frac{\alpha\sqrt{a_{12}a_{21}}}{\sqrt{\alpha^2a_{12}a_{21}+4a_{12}a_{21}}}   \Bigg)\\
    &=\frac{1}{2}\Bigg(1-\frac{\alpha\sqrt{a_{12}a_{21}}}{\sqrt{\alpha^2+4}\sqrt{a_{12}a_{21}}}   \Bigg)
    =\frac{1}{2}\Bigg(1-\frac{\alpha}{\sqrt{\alpha^2+4}}\Bigg).
\end{aligned}    
\]
Analogously
\[
    \lambda'_{max}(a_{22}+\alpha\sqrt{a_{12}a_{21}})=\frac{1}{2}\Bigg(1+\frac{\alpha}{\sqrt{\alpha^2+4}}\Bigg).
\]
Finally
\[
    \lambda'_{max}(2a_{22})=\frac{1}{2}\Bigg(1+\frac{a_{22}}{\sqrt{a_{22}^2+4a_{12}a_{21}}}\Bigg).
\]
Observe now that 
\[
\begin{aligned}
    \lambda'_{max}(a_{22}-\alpha\sqrt{a_{12}a_{21}})-\lambda'_{max}(0)&=\frac{1}{2}\Bigg(1-\frac{\alpha}{\sqrt{\alpha^2+4}}\Bigg)-\frac{1}{2}\Bigg(1-\frac{a_{22}}{\sqrt{a_{22}^2+4a_{12}a_{21}}} \Bigg)\\
    &-\frac{\alpha}{2\sqrt{\alpha^2+4}}+\frac{a_{22}}{2\sqrt{a_{22}^2+4a_{12}a_{21}}}\\
    &=\lambda'_{max}(2a_{22})-\lambda'_{max}(a_{22}+\alpha\sqrt{a_{12}a_{21}}).
\end{aligned}
\]
Also note that 
\[
    \lambda'_{max}(a_{22}+\alpha\sqrt{a_{12}a_{21}})-\lambda'_{max}(a_{22}-\alpha\sqrt{a_{12}a_{21}})=\frac{\alpha}{\sqrt{4+\alpha^2}}.
\]
Then
\[
\begin{aligned}
    &\frac{\lambda'_{max}(a_{22}+\alpha\sqrt{a_{12}a_{21}})-\lambda'_{max}(a_{22}-\alpha\sqrt{a_{12}a_{21}})}{\lambda'_{max}(a_{22}-\alpha\sqrt{a_{12}a_{21}})-\lambda'_{max}(0)}\\
    &=\frac{\lambda'_{max}(a_{22}+\alpha\sqrt{a_{12}a_{21}})-\lambda'_{max}(a_{22}-\alpha\sqrt{a_{12}a_{21}})}{\lambda'_{max}(2a_{22})- \lambda'_{max}(a_{22}+\alpha\sqrt{a_{12}a_{21}})}
    =\frac{\frac{\alpha}{\sqrt{4+\alpha^2}}}{-\frac{\alpha}{2\sqrt{4+\alpha^2}}+\frac{a_{22}}{2\sqrt{a_{22}^2+4a_{12}a_{21}}}}\\
    &=\frac{2\alpha \sqrt{a_{22}^2+4a_{12}a_{21}}}{a_{22}\sqrt{\alpha^2+4}-\alpha\sqrt{a_{22}^2+4a_{12}a_{21}}}\geq \frac{2\alpha}{\sqrt{\alpha^2+4}-\alpha}.
\end{aligned}    
\]
Since $\sqrt{a_{22}^2+4a_{12}a_{21}}> a_{22}$ and
\[
    a_{22}\sqrt{\alpha^2+4}-\alpha\sqrt{a_{22}^2+4a_{12}a_{21}}\leq a_{22}\sqrt{\alpha^2+4}-\alpha a_{22}
\]
and the claim follows.
\end{proof}

\section{Proof of Corollary \ref{cor:relation_H}}
\label{appendix_C}
\begin{proof}
We show the result for $\Vec{H}$, which is defined in \eqref{def: H}. Then, the result for $\Vec{I}$ follows straightforwardly using \eqref{eq:relation_H_I}.\\
Note that $\lambda_2>\lambda_1>\lambda_0$ and $\tilde{\lambda_2} > \tilde{\lambda_1}> \tilde{\lambda_0}$. 
We proceed componentwise. Since $V^0_1>0, V^2_1>0$,
\[
\begin{aligned}
    |{H}_1^{\lambda_2}(t)-{H}_1^{\lambda_0}(t)|&=|e^{\lambda_2 t}W^2_1+e^{\widetilde{\lambda_2}t}V^2_1-e^{\lambda_0 t}W^0_1-e^{\widetilde{\lambda_0}t}V^0_1|\\
    &>e^{\lambda_2 t}W^2_1 \Bigg|1-\frac{W^0_1}{W^2_1}e^{(\lambda_0-\lambda_2)t}-\frac{V^0_1}{W^2_1}e^{(\widetilde{\lambda_0}-\lambda_2)t}\Bigg|
\end{aligned}    
\]
and noticing that
\[
\begin{aligned}
\lambda_0-\lambda_2&<-\sqrt{2}\sqrt{a_{21}a_{12}}\\
\widetilde{\lambda_0}-\lambda_2&<\widetilde{\lambda_2}-\lambda_2=-2\sqrt{2}\sqrt{a_{21}a_{12}}
\end{aligned}
\]
we get
\[
    |{H}^{\lambda_2}_1(t)-{H}^{\lambda_0}_1(t)|>e^{\lambda_2 t}W^2_1 \Bigg|1-\bigg(\frac{W^0_1}{W^2_1}+\frac{V^0_1}{W^2_1}\bigg) e^{-\sqrt{2a_{12}a_{21}}t}\Bigg|
\]
and setting $\frac{W^0_1+V^0_1}{W^2_1}=:Q_1$ we can pick up $t$ such that $1-Q_1e^{-\sqrt{2a_{12}a_{21}}t}>\frac{1}{2}$ that is $e^{\sqrt{2a_{12}a_{21}}t}>2 Q_1$ hence $t>\frac{1}{\sqrt{2a_{12}a_{21}}}\ln(2Q_1)=:t^0_1$ so that finally
\begin{equation}\label{lowerbound}
|{H}^{\lambda_2}_1(t)-{H}^{\lambda_0}_1(t)|>\frac{1}{2}W_1e^{\lambda_2t}.
\end{equation}
On the other hand 
\begin{equation}\label{eq:inequalityI1}
\begin{aligned}
    k|H^{\lambda_1}_1(t)-H^{\lambda_0}_1(t)|&=k|e^{\lambda_1 t}W^1_1+e^{\widetilde{\lambda_1}t}V^1_1-e^{\lambda_0 t}W^0_1-e^{\widetilde{\lambda_0}t}V^0_1|\\
    &\leq k e^{\lambda_1t}W^1_1\bigg(1+e^{(\widetilde{\lambda_1}-\lambda_1)}\frac{V^1_1}{W^1_1}+e^{(\lambda_0-\lambda_1)t}\frac{W^0_1}{W^1_1}+e^{(\widetilde{\lambda_0}-\lambda_1)t}\frac{V^0_1}{W^1_1}\bigg).
\end{aligned}    
\end{equation}
Using the fact that $\widetilde{\lambda_1}-\lambda_1<0$, $\lambda_0-\lambda_1<0$ and $\widetilde{\lambda_0}-\lambda_1<\widetilde{\lambda_1}-\lambda_1<0$, from \eqref{eq:inequalityI1} we get
\begin{equation}\label{upperbound}
    k|H^{\lambda_1}_1(t)-H^{\lambda_0}_1(t)| \leq k e^{\lambda_1t}W^1_1\bigg(1+\frac{V^1_1+W^0_1+V^0_1}{W^1_1}\bigg)=k W_1 e^{\lambda_1 t}\bigg(1+Q_1+\frac{V^1_1}{W^1_1}\bigg).
\end{equation}
Hence, by \eqref{lowerbound} and \eqref{upperbound} we then get 
\[
    \frac{1}{2}W_1e^{\lambda_2t}> k W_1 e^{\lambda_1 t}\bigg(1+Q_1+\frac{V^1_1}{W^1_1}\bigg)
\]
that is 
\[
    e^{(\lambda_2-\lambda_1)t}>2k\bigg(1+Q_1+\frac{V^1_1}{W^1_1}\bigg)
\]
\[
    e^{2\sqrt{a_{12}a_{21}}t}>2k\bigg(1+Q_1+\frac{V^1_1}{W^1_1}\bigg)
\]
hence
\[
    t>\frac{1}{2\sqrt{a_{12}a_{21}}}\ln\bigg(2k\bigg(1+Q_1+\frac{V^1_1}{W^1_1}\bigg)\bigg)=:t^1_1.
\]
Therefore, for $t\geq \textrm{max}(t^0_1,t^1_1)$
\[
    |H^{\lambda_2}_1(t)-H^{\lambda_0}_1(t)|>k|H^{\lambda_1}_1(t)-H^{\lambda_0}_1(t)|.
\]
Analogously one can show that 
\[
    |H^{\lambda_2}_2(t)-H^{\lambda_0}_2(t)|\geq k|H^{\lambda_1}_2(t)-H^{\lambda_0}_2(t)|
\]
for $t\geq \textrm{max}(t^0_2,t^1_2)$, where 
\[
\begin{aligned}
    t^0_2&=\frac{1}{\sqrt{2a_{21}a_{12}}}\ln(2Q_2),\\
    Q_2&=\frac{W^0_2+V^0_2}{W^2_2},\\
    t^1_2&=\frac{1}{2\sqrt{a_{12}a_{21}}}\ln\bigg(2k\bigg(1+Q_2+\frac{V^2_2}{W^2_2}\bigg)\bigg).
\end{aligned}    
\]
So picking up $t>\textrm{max}(t^0_1,t_1^1,t^0_2,t^1_2)$ and using \eqref{eq:relation_H_I}, the claim follows.
\end{proof}

\section{An application of Theorem \ref{EigenvaluesThm} }
\label{Consequencies}
Inequality \eqref{eq:eigenvalues_ineq} indicates a \textit{phase transition}. In fact from now on let $a_{22}^2>>a_{12}a_{21}$ and $\alpha=2$. In this case 
\[
    \frac{2\alpha}{\sqrt{\alpha^2+4}-\alpha}\simeq 4.8
\]
and so by \eqref{eq:eigenvalues_ineq} we have
\[
    \Delta_1\geq 4.8 \Delta_0=4.8 \Delta_2.
\]
Given
\[
    \lambda'_{max}(0)=\frac{1}{2}\Bigg(1-\frac{a_{22}}{\sqrt{a_{22}^2+4a_{12}a_{21}}}\Bigg)
\]
and
\[
    \lambda'_{max}(a_{22}-\alpha\sqrt{a_{12}a_{21}})=\frac{1}{2}\Bigg(1-\frac{\alpha}{\sqrt{\alpha^2+4}}  \Bigg)
\]
if $\lambda'_{max}(0)<<\lambda'_{max}(a_{22}-\alpha\sqrt{a_{12}a_{21}})/4.8$ then from \eqref{eq:eigenvalues_ineq}
\begin{equation}\label{boundder}
\begin{aligned}
    \lambda'_{max}(a_{22}+\alpha\sqrt{a_{12}a_{21}})&\geq 5.8 \lambda'_{max}(a_{22}-\alpha\sqrt{a_{12}a_{21}})-4.8 \lambda'_{max}(0)\\
    &\geq 4.8 \lambda'_{max}(a_{22}-\alpha\sqrt{a_{12}a_{21}}).
\end{aligned}    
\end{equation}
Hence, by \eqref{boundder} and since $\lambda'_{max}$ is increasing
\[
\begin{aligned}
    4.8\Bigg(\lambda_{max}(a_{22}-\alpha\sqrt{a_{12}a_{21}})-\lambda_{max}(0)\Bigg)&\leq 4.8 \lambda'_{max}(a_{22}-\alpha \sqrt{a_{21}a_{12}})(a_{22}-\alpha\sqrt{a_{12}a_{21}})\\
    &\leq \lambda'_{max}(a_{22}+\alpha\sqrt{a_{12}a_{21}}) (2a_{22}-(a_{22}+\alpha\sqrt{a_{12}a_{21}}))\\
    &\leq \lambda_{max}(2a_{22})-\lambda_{max}(a_{22}+\alpha\sqrt{a_{21}a_{12}}).
\end{aligned}    
\]

Therefore

\[
    \lambda_{max}(2a_{22})-\lambda_{max}(0)\geq 4.8(\lambda_{max}(a_{22}-\alpha\sqrt{a_{21}a_{12}})-\lambda_{max}(0)).
\]
Hence $\forall 0<a_{11}\leq a_{22}-\alpha\sqrt{a_{12}a_{21}}$, we have
\[
    \lambda_{max}(2a_{22})-\lambda_{max}(0)\geq 4.8(\lambda_{max}(a_{11})-\lambda_{max}(0)).
\]
Last inequality tells us that closing schools starting from a reproduction number $2a_{22}$ is much more effective than starting from a reproduction number around $a_{22}$.

\section{Proof of Theorem \ref{attack-rates}}
\label{appendix_E}

\begin{proof}
Let 
\begin{eqnarray}
\widetilde{\mathcal{I}}_1^{(k)} &:=&\int_0^{+\infty}I_1(t) e^{-k v t}dt, \\
\widetilde{\mathcal{I}}_2^{(k)} &:=&\int_0^{+\infty}I_2(t) e^{-k v t} dt.
\end{eqnarray}
Integrating the first equation in 
\eqref{eq:sir_model_vacc_linear extract},
we have
\begin{equation}
      -I_1(0) = (\widetilde{\beta}_{11}S_1(0)-1)\widetilde{\mathcal{I}}_1^{(0)}+\widetilde{\beta}_{12}S_1(0) \widetilde{\mathcal{I}}_2^{(0)}
      \end{equation}
      which implies
\begin{equation}\label{J1}  \widetilde{\mathcal{I}}_1^{(0)}=\frac{\widetilde{\beta}_{12}S_1(0)\widetilde{\mathcal{I}}_2^{(0)}+I_1(0)}{1-\widetilde{\beta}_{11}S_1(0)} > 0\,\,\, \text{   if   }\,\, \widetilde{\beta}_{11}S_1(0) < 1.  
\end{equation}
This proves \eqref{Jvalue} (note that $\widetilde{\mathcal{I}}_1^{(0)}\equiv \widetilde{\mathcal{I}}_1$).

Integrating the second equation in 
\eqref{eq:sir_model_vacc_linear extract}
we get
\begin{equation}\label{intSecondEq}
     -I_2(0) = \widetilde{\beta}_{21}S_2(0)\widetilde{\mathcal{I}}_1^{(1)} + \widetilde{\beta}_{22}S_2(0)\widetilde{\mathcal{I}}_2^{(1)} - \widetilde{\mathcal{I}}_2^{(0)} 
\end{equation}
which implies
\begin{equation}\label{H(J,H_1)}
      \widetilde{\mathcal{I}}_2^{(0)} = \widetilde{\mathcal{I}}_2^{(0)}(\widetilde{\mathcal{I}}_1^{(1)},\widetilde{\mathcal{I}}_2^{(1)}) = \widetilde{\beta}_{21}S_2(0)\widetilde{\mathcal{I}}_1^{(1)} + \widetilde{\beta}_{22}S_2(0)\widetilde{\mathcal{I}}_2^{(1)} + I_2(0).
\end{equation}
Substituting in \eqref{J1}, we get
\begin{equation} \label{J(J,H_1)}
    \widetilde{\mathcal{I}}_1^{(0)}=\widetilde{\mathcal{I}}_1^{(0)}(\widetilde{\mathcal{I}}_1^{(1)},\widetilde{\mathcal{I}}_2^{(1)})=
    \frac{\widetilde{\beta}_{12}S_1(0)(\widetilde{\beta}_{21}S_2(0)\widetilde{\mathcal{I}}_1^{(1)} + \widetilde{\beta}_{22}S_2(0)\widetilde{\mathcal{I}}_2^{(1)} + I_2(0))+I_1(0)}{1-\widetilde{\beta}_{11}S_1(0)}
\end{equation}
which in turn implies
\begin{equation}\label{H(H_1)}
\begin{aligned}
    \widetilde{\mathcal{I}}_2^{(0)}&=\widetilde{\mathcal{I}}_2^{(0)}(\widetilde{\mathcal{I}}_2^{(1)})\\
    &=\widetilde{\beta}_{21}S_2(0)\widetilde{\mathcal{I}}_1^{(1)}+\widetilde{\beta}_{22}S_2(0)\widetilde{\mathcal{I}}_2^{(1)}+I_2(0) \\
    &=\widetilde{\beta}_{21}S_2(0)\frac{\frac{I_1(0)}{v}+\frac{1}{v}\widetilde{\beta}_{12}S_1(0)\widetilde{\mathcal{I}}_2^{(1)}}{1-\frac{1}{v}(\widetilde{\beta}_{11}S_1(0)-1)}+\widetilde{\beta}_{22}S_2(0)\widetilde{\mathcal{I}}_2^{(1)}+I_2(0) \\
    &=\frac{\widetilde{\beta}_{21}S_2(0)I_1(0)}{v-(\widetilde{\beta}_{11}S_1(0)-1)}+I_2(0)+\widetilde{\mathcal{I}}_2^{(1)}\left(\widetilde{\beta}_{22}S_2(0)+\frac{\widetilde{\beta}_{21}S_2(0)\widetilde{\beta}_{12}S_1(0)}{v-(\widetilde{\beta}_{11}S_1(0)-1)}\right) \\
    &= p_1 + \widetilde{\mathcal{I}}_2^{(1)} q_1.
\end{aligned}    
\end{equation}
For each $k \geq 1$, it follows from    
\eqref{eq:sir_model_vacc_linear extract}, 
and integration by parts that
\begin{equation}\label{J}
\begin{aligned}
    \widetilde{\mathcal{I}}_1^{(k)} &= \int^{+\infty}_0 I_1 e^{-k v t} dt\\ 
    &=\frac{I_1(0)}{k v} + \int^{+\infty}_0(\widetilde{\beta}_{11}S_1(0)I_{1}+\widetilde{\beta}_{12}S_1(0)I_{2}-I_1)\frac{e^{-k vt}}{k v} dt \\
    &=\frac{I_1(0)}{k v}+\frac{1}{k v}\left((\widetilde{\beta}_{11}S_1(0)-1)\widetilde{\mathcal{I}}_1^{(k)}+\widetilde{\beta}_{12}S_1(0)\widetilde{\mathcal{I}}_2^{(k)}\right)
\end{aligned}
\end{equation}
which yields
\begin{equation} \label{J(H_1)}
\begin{aligned}
    \widetilde{\mathcal{I}}_1^{(k)}(\widetilde{\mathcal{I}}_2^{(k)}) = \frac{I_1(0)+\widetilde{\beta}_{12}S_1(0)\widetilde{\mathcal{I}}_2^{(k)}}{k v-(\widetilde{\beta}_{11}S_1(0)-1)} > 0,
\end{aligned} 
\end{equation}
and 
\[ 
\begin{aligned}
    \widetilde{\mathcal{I}}_2^{(k)} &= \int^{+\infty}_{0} I_2 e^{-k vt} dt =\\
    &= \frac{I_2(0)}{k v}+\int^{+\infty}_{0} \left(\widetilde{\beta}_{21}S_1(0)e^{-vt}I_1 + \widetilde{\beta}_{22}S_1(0)e^{- vt}I_2 - I_2\right)\frac{e^{-kvt}}{kv} dt\\
    &= \frac{1}{kv}\left(I_2(0) +\widetilde{\beta}_{21}S_2(0) \widetilde{\mathcal{I}}_1^{(k+1)}
    +\widetilde{\beta}_{22}S_2(0)\widetilde{\mathcal{I}}_2^{(k+1)}- \widetilde{\mathcal{I}}_2^{(k)}\right)
\end{aligned}
\]
that gives
\[
    \widetilde{\mathcal{I}}_2^{(k)} = \frac{1}{(kv+1)}
    (I_2(0)+\widetilde{\beta}_{21}S_1(0)\widetilde{\mathcal{I}}_1^{(k+1)}+\widetilde{\beta}_{22}S_2(0)\widetilde{\mathcal{I}}_2^{(k+1)}).
\]
This implies
\begin{equation} \label{H2k}
\begin{aligned}
    \widetilde{\mathcal{I}}_2^{(k)}(\widetilde{\mathcal{I}}_2^{(k+1)})&=\frac{1}{kv+1}\left(I_2(0)+\widetilde{\beta}_{21}S_2(0)\frac{I_1(0)+\widetilde{\beta}_{12}S_1(0)\widetilde{\mathcal{I}}_2^{(k+1)}}{(k+1)v-\widetilde{\beta}_{11}S_1(0)+1}
    +\widetilde{\beta}_{22} S_2(0) \widetilde{\mathcal{I}}_2^{(k+1)}\right) \\
    &=\frac{1}{kv+1}  \left(    I_2(0)+\frac{\widetilde{\beta}_{21}S_2(0)I_1(0)    }{(k+1)v-\widetilde{\beta}_{11}S_1(0)+1} \right. \\
    & \left.
    \hskip 2cm + \widetilde{\mathcal{I}}_2^{(k+1)}\left(\widetilde{\beta}_{22} S_2(0)+ \frac{\widetilde{\beta}_{12}\widetilde{\beta}_{21}S_1(0)S_2(0)}{(k+1)v-\widetilde{\beta}_{11}S_1(0)+1}\right) \right) \\
    & =\frac{1}{kv+1} (p_{k+1}+ \widetilde{\mathcal{I}}_2^{(k+1)} q_{k+1}  ).
\end{aligned}
\end{equation}
From \eqref{H(H_1)} and \eqref{H2k}
\[
\begin{aligned}
    \widetilde{\mathcal{I}}_2^{(0)} &= p_1 + q_1\widetilde{\mathcal{I}}_2^{(1)}\\
    &= p_1 + q_1 \frac{1}{v+1}(p_2+q_2H_2)\\
    &=...\\
    &=  \sum_{i=1}^{\infty} p_i(\prod_{r=1}^{i-1}
\frac{q_r}{r v+1}).
\end{aligned}
\]
This proves \eqref{Hvalue} (note that $\widetilde{\mathcal{I}}_2^{(0)}\equiv \widetilde{\mathcal{I}}_2$). {Using Wolfram Mathematica for symbolic calculations}, we provide an explicit expression for $\widetilde{\mathcal{I}}_2^{(0)}(\widetilde{\beta}_{11})$ in terms of special functions
\[
\begin{aligned}
    &\Big\{S_2(0) (1 - \widetilde{\beta}_{11} S_1(0) + 
   v)\Big\}^{-1}\Big\{\big[I_2(0)S_2(0)\big(1-\widetilde{\beta}_{11}S_1(0)+v\big)\big]\times \\
   &\times\  _2F_2\Big(1, 
     1 + \frac{1}{v} - \frac{\widetilde{\beta}_{11} S_1(0)}{v} + \frac{\widetilde{\beta}_{12} \widetilde{\beta}_{21} S_1(0)}{v\widetilde{\beta}_{22}}; 1 + 
      \frac{1}{v}, 1 + \frac{1}{v} - \frac{\widetilde{\beta}_{11} S_1(0)}{v}; \frac{\widetilde{\beta}_{22}S_2(0)}{v}\Big)\\
     &  +\widetilde{\beta}_{21} I_1(0) S^2_2(0)\  _2F_2\Big(1, 
     1 + \frac{1}{v} - \frac{\widetilde{\beta}_{11} S_1(0)}{v} + \frac{\widetilde{\beta}_{12} \widetilde{\beta}_{21} S_1(0)}{v \widetilde{\beta}_{22}}; 1 + 
      \frac{1}{v}, 2 + \frac{1}{v} - \frac{\widetilde{\beta}_{11} S_1(0)}{v}; \frac{\widetilde{\beta}_{22} S_2(0)}{v}\Big)\Big\}
\end{aligned}      
\]
where $\ _pF_q(\Vec{a};\Vec{b};z)$ is the generalized hypergeometric function, see for example \cite{Erdelyi et al}.

\end{proof}

\vskip 2cm

Contact address: New York University Abu Dhabi, Division of Science, Abu Dhabi, 129188, UAE

Contact e-mail: ag189@nyu.edu


\vskip 0.5cm
\bigskip






\small 
\begin{thebibliography}{99}

\bibitem{Byrne}  Byrne, A. et al. 
Inferred duration of infectious period of SARS-CoV-2: rapid scoping review and analysis of available evidence for asymptomatic and symptomatic COVID-19 cases. 
\textit{BMJ Open} \textbf{10}, (2020).


\bibitem{Bertozzi et al} Bertozzi A. et al. The challenges of modeling and forecasting the spread of COVID-19, \textit{PNAS} July 21, 2020 117 (29) 16732-16738;

\bibitem{Brauner et al} Brauner et al. Inferring the effectiveness of government interventions against COVID-19, \textit{Science}, 19 Feb 2021:
Vol. 371, Issue 6531, DOI: 10.1126/science.abd9338

\bibitem{Castro et al.} Castro M., Ares S., Cuesta J. A., Manrubia S., The turning point and end of an expanding epidemic cannot be precisely forecast. \textit{Proceedings of the National Academy of Sciences} Oct 2020, \textbf{117} (42) 26190-26196.


\bibitem{Gandini et al} Gandini et al. A cross-sectional and prospective cohort study of the role of schools in the SARS-CoV-2 second wave in Italy,\textit{The Lancet Regional Health - Europe}, VOLUME \textbf{5}, 100092, (2021)

\bibitem{Gold et al.}Gold JA, Gettings JR, Kimball A, et al. Clusters of SARS-CoV-2 Infection Among Elementary School Educators and Students in One School District — Georgia, December 2020–January 2021, MMWR Morb Mortal Wkly Rep 2021;70:289–292

\bibitem{Roda et al.} Roda W., Varughese M.B., Han D.,and  Lia M., Why is it difficult to accurately predict the COVID-19 epidemic? \textit{Infect Dis Model.} 2020; \textbf{5}: 271–281. Published online 2020 Mar 25. doi: 10.1016/j.idm.2020.03.001




 \bibitem{Erdelyi et al} Erdélyi A., Magnus W.,  Oberhettinger F. and Tricomi F., Higher Transcendental Functions Volume I, McGraw-Hill Book Company; 1st Edition (November 1, 1953.
 
\bibitem{Rozhnova et. al} Rozhnova G. et al. Model-based evaluation of school- and non-school-related measures to control the COVID-19 pandemic,
\textit{Nature Communications} volume \textbf{12}, Article number: 1614 (2021)

\bibitem{PCJ17}
Prem K. et al. 
Projecting social contact matrices in 152 countries using contact surveys and demographic data. 
\textit{PLOS Computational Biology} \textbf{13}, (2017).

\bibitem{gandolfiI_2(0)} Gandolfi, A.
Planning of school teaching during Covid-19. 
\textit{Physica D: Nonlinear Phenomena} \textbf{415}, (2021).

\bibitem{De la Sen et. al}
De la Sen, M., Ibeas, A. On an SE(Is)(Ih)AR epidemic model with combined vaccination and antiviral controls for COVID-19 pandemic. Adv Differ Equ 2021, 92 (2021). https://doi.org/10.1186/s13662-021-03248-5

\bibitem{Lee et al.} Lee, B. et al.
Modeling the impact of school reopening on SARS-CoV-2 transmission using contact structure data from Shanghai.
\textit{BMC Public Health} \textbf{20}, (2020). 

\bibitem{Viner2020} Viner, R. M. et al. 
School closure and management practices during coronavirus outbreaks including COVID-19: a rapid systematic review.
\textit{The Lancet Child \& Adolescent Health} \textbf{4}, 397-404 (2020).

\bibitem{Walsh et al. 21}  Walsh, S. et al. 
Do school closures reduce community transmission of COVID-19? A systematic review of observational studies. 
Preprint at \url{https://www.medrxiv.org/content/10.1101/2021.01.02.21249146v1.full-text} 
(2021).

\bibitem{Cao2020} Cao, W. et al. 
The psychological impact of the COVID-19 epidemic on college students in China.
\textit{Psychiatry Research} \textbf{287}, (2020).

\bibitem{Libotte et al.20} Libotte, G. B. et al. 
Determination of an optimal control strategy for vaccine administration in COVID-19 pandemic treatment. 
\textit{Computer Methods and Programs in Biomedicine} \textbf{196}, (2020).

\bibitem{Singh2020} Singh, S. et al. 
Impact of COVID-19 and lockdown on mental health of children and adolescents: A narrative review with recommendations. 
\textit{Psychiatry Research} \textbf{293} (2020).

\bibitem{GriffithS_2(0)20} Panovska-Griffiths, J. et al. 
Determining the optimal strategy for reopening schools,the impact of test and trace interventions, and the risk of occurrence of a second COVID-19 epidemic wave in the UK: a modelling study.
\textit{The Lancet Child \& Adolescent Health} \textbf{4}, 817-827 (2020).

\bibitem{Kim2020} Soyoung, K. et al. 
School opening delay effect on transmission dynamics of Coronavirus disease 2019 in Korea: Based on mathematical modeling and simulation study. 
\textit{Journal of Korean Medical Science} \textbf{35}, (2020).

\bibitem{WHO} What we know about
COVID-19 transmission in schools: the latest on the COVID-19 global situation \& the spread of COVID-19 in schools. 
\textit{World Health Organization Report}, \url{https://www.who.int/docs/default-source/coronaviruse/risk-comms-updates/update39-covid-and-schools.pdf?sfvrsn=320db233_2} (October 2020). 

\bibitem{Brauer2017} Brauer, F. 
Mathematical Epidemiology: Past, present, and future. 
\textit{Infectious Disease Modelling} \textbf{2}, 113-127 (2017).

\bibitem{Keeling2008} Keeling, M. J. et al. 
Modelling infectious diseases in humans and animals.
\textit{Princeton University Press} Chapter 3, (2008).

\bibitem{Iwata2020} Iwata, K. et al. 
Was school closure effective in mitigating coronavirus disease 2019 (COVID-19)? Time series analysis using Bayesian inference.
\textit{International Journal of Infectious Diseases} \textbf{99}, 57-61 (2020).

\bibitem{Matzinger2020} Matzinger, P. et al. 
Strong impact of closing schools, closing bars and wearing masks during the COVID-19 pandemic: results from a simple and revealing analysis.
Preprint at \url{https://www.medrxiv.org/content/10.1101/2020.09.26.20202457v1.full-text} (2020).

\bibitem{Ismail2020} Ismail, S. A. et al. 
SARS-CoV-2 infection and transmission in educational settings: a prospective, cross-sectional analysis of infection clusters and outbreaks in England.
\textit{The Lancet Infectious Diseases} \textbf{21}, 344-353 (2021).

\bibitem{Larosa2020} Larosa, E. et al. 
Secondary transmission of COVID-19 in preschool and school settings after their reopening in northern Italy: a population-based study.
Preprint at \url{https://www.medrxiv.org/content/10.1101/2020.11.17.20229583v1.full-text} (2020).

\bibitem{Flaxman2020} Flaxman, S. et al. 
Estimating the effects of non-pharmaceutical interventions on COVID-19 in Europe.
\textit{Nature} \textbf{584} 257–261 (2020).

\bibitem{Singanayagam2020} Singanayagam, A. et al. 
Duration of infectiousness and correlation with RT-PCR cycle threshold values in cases of COVID-19, England, January to May 2020.
\textit{Euro Surveill} \textbf{25}, (2020).  
 
\bibitem{Blyuss2021} Blyuss, K. B. et al.
Effects of Latency and Age Structure on the Dynamics and Containment of COVID-19.
\textit{Journal of Theoretical Biology} \textbf{513} (2021).

\bibitem{Yuan2021} Yuan, P. et al. 
School and community reopening during the COVID-19 pandemic: a mathematical modeling study.
Preprint at \url{https://www.medrxiv.org/content/10.1101/2021.01.13.21249753v1.full} (2021).

\bibitem{Domenico2021} Di Domenico, L. et al. 
Modelling safe protocols for reopening schools during the COVID-19 pandemic in France. 
\textit{Nature Communications} \textbf{12} (2021).

\bibitem{Espana2021} España, G. et al. 
The impact of school reopening on COVID-19 dynamics in
Bogotá, Colombia. 
Preprint at \url{https://osf.io/ebjx9/} (2021).

\bibitem{Keskinocak et al} Keskinocak, P., Asplund, J., Serban, N. \& Oruc Aglar, B. E. Evaluating scenarios for school reopening under COVID19. Preprint at medRxiv https://doi.org/10.1101/2020.07.22.20160036 (2020).


\bibitem{Voinsky2020} Voinsky, I. et al. 
Effects of age and sex on recovery from  COVID-19:  Analysis  of  5769  Israeli  patients. 
\textit{The  Journal  of  Infection} \textbf{81}, (2020).

\bibitem{Oran2020} Oran, D.P., Topol, E.J. 
Prevalence of Asymptomatic SARS-CoV-2 Infection. \textit{Annals of Internal Medicine} \textbf{173}, (2020).

\bibitem{Davies2020} Davies, N.G. et al.
Age-dependent effects in the transmission and control of COVID-19 epidemics. \textit{Nature Medicine} \textbf{26}, (2020).

\bibitem{Dong2020} Dong, Y. et al
Epidemiology of COVID-19 Among Children in China. \textit{Pediatrics} \textbf{145}, (2020).
\end{thebibliography}
\end{document}